\pdfoutput=1
\RequirePackage{ifpdf}
\ifpdf 
\documentclass[pdftex]{sigma}
\else
\documentclass{sigma}
\fi

\numberwithin{equation}{section}
\DeclareMathOperator{\tr}{tr}

\def\bra#1{\left\langle #1\right|}

\def\ket#1{\left| #1\right\rangle}

\newcommand{\cg}[6]{
 \left(
 \begin{array}{@{}cc|c@{}}
 #1 & #3 & #5
 \\
 #2 & #4 & #6
 \end{array}
 \right)
}

\begin{document}

\newcommand{\arXivNumber}{1403.0808}

\allowdisplaybreaks

\renewcommand{\thefootnote}{$\star$}

\renewcommand{\PaperNumber}{086}

\FirstPageHeading

\ShortArticleName{Matrix Bases for Star Products: a~Review}

\ArticleName{Matrix Bases for Star Products: a~Review\footnote{This paper is a~contribution to the Special Issue on Deformations of Space-Time
and its Symmetries.
The full collection is available at \href{http://www.emis.de/journals/SIGMA/space-time.html}
{http://www.emis.de/journals/SIGMA/space-time.html}}}

\Author{Fedele LIZZI~$^{\dag\ddag\S}$ and Patrizia VITALE~$^{\dag\ddag}$}

\AuthorNameForHeading{F.~Lizzi and P.~Vitale}

\Address{$^\dag$~Dipartimento di Fisica, Universit\`{a} di Napoli Federico II, Napoli, Italy}
\EmailD{\href{mailto:email@address}{fedele.lizzi@na.infn.it}, \href{mailto:email@address}{patrizia.vitale@na.infn.it}}

\Address{$^\ddag$~INFN, Sezione di Napoli, Italy} 
\Address{$^\S$~Institut de Ci\'encies del Cosmos, Universitat de Barcelona, Catalonia, Spain}

\ArticleDates{Received March 04, 2014, in f\/inal form August 11, 2014; Published online August 15, 2014}

\Abstract{We review the matrix bases for a~family of noncommutative $\star$ products based on a~Weyl map.
These products include the Moyal product, as well as the Wick--Voros products  
and other translation invariant ones.
We also review the derivation of Lie algebra type star products, with adapted matrix bases.
We discuss the uses of these matrix bases for f\/ield theory, fuzzy spaces and emergent gravity.}


\Classification{58Bxx; 40C05; 46L65}

\renewcommand{\thefootnote}{\arabic{footnote}}
\setcounter{footnote}{0}

\section{Introduction}

Star products were originally introduced~\cite{Groenewold, Moyal} in the context of point particle quantization, they
were generalized in~\cite{BFFLS1,BFFLS2} to comprise general cases, the physical motivations behind this was the
quantization of phase space.
Later star products became a~tool to describe possible noncommutative geometries of spacetime itself (for a~review
see~\cite{WSS}).
In this sense they have become a~tool for the study of f\/ield theories on noncommutative spaces~\cite{Szabo}.
In this short review we want to concentrate on one particular aspect of the star product, the fact that any action
involving f\/ields which are multiplied with a~star product can be seen as a~matrix model.
This is because the noncommutative star products algebras can always be represented as operators on a~suitable Hilbert
space, and one can then just consider the matrix representation of these operators.
The matrix representation is not only a~useful tool, but also of\/fers a~conceptual interpretation of noncommutative
spaces.
It shows that, in analogy with quantum phase space, the proper setting of deformed products is noncommutative
geometry~\cite{Connes}, seen as a~spectral theory of operators.

This is a~partial review, as we do not have the ambition to cover exhaustively all known products, nor all known bases.
We will consider products which are connected to symmetries and which have developments in f\/ield theory.
Even then we will not cover completely all work done, and apologize for the omissions.

In order to construct the matrix bases we will use mainly quantization-dequantization maps, whereby we associate
operators to functions, and viceversa.
We will f\/irst consider a~family of translation-invariant star products induced on suitable algebras of functions on
${\mathbb R}^{2n}$ through $s$-ordered quantization-dequantization maps.
In this approach noncommutative algebras are obtained as symbols of algebras of operators, which are def\/ined in terms of
a~special family of operators using the trace formula (what is sometimes called the `dequantization' map because of its
original meaning in the Wigner--Weyl formalism), while the reconstruction of operators in terms of their symbols (the
`quantization' map) is determined using another family of operators.
These two families determine completely the noncommutative algebra, including the kernel of the star-product.

Once we have established the connection, in Section~\ref{Section4} we construct matrix bases for $s$-ordered products and for the Moyal
and Wick--Voros cases in more detail.
In Section~\ref{Section5} we review the construction of a~family of star products of Lie algebra type, obtained by reduction of
higher-dimensional, translation invariant ones.
We thus discuss in Section~\ref{Section6} the matrix basis for one of them on the space ${\mathbb R}^3$. Section~\ref{Section7}
is devoted to the application to quantum f\/ield theory of the matrix bases introduced previously.
Finally, Section~\ref{Section8} is devoted to fuzzy spaces.

\section{The star product}\label{Section2}

In this section we review a~general construction for the noncommutative algebra of operator symbols acting on a~Hilbert
space $\mathcal H$.
We follow the presentation and the notation of~\cite{marmo2, MMV04}.

Given an operator $\hat A$ acting on the Hilbert space $\mathcal H$ (which can be f\/inite- or inf\/inite-dimen\-si\-onal), let
us have two distinct operator bases for the operators acting on ${\mathcal H}$, $\hat U(x)$ and $\hat D (x)$.
The two bases are labelled by a~set of parameters $x=(x_1,\dots,x_N)$,
with $x_k\in \mathbb K$, $k=1, \dots ,N$,
with $\mathbb K$ any f\/ield, usually the reals, complex or integers.
The symbol $f_A(x)$ of the operator $\hat A$ is the following function of $x$:
\begin{gather}
f_A=\Omega^{-1}\big(\hat A\big)\equiv \tr \big(\hat A\hat U(x)\big).
\label{deq-map}
\end{gather}
We assume that the trace exists for all parameters~$x$, although often the function~$f$ may be a~distribution.
If we now consider the ``parameters''~$x$ as the coordinates of a~manifold it makes sense to def\/ine the inverse map (the
``reconstruction formula''),
which associates operators to functions, is def\/ined in terms of the second family of
operators, that is
\begin{gather}
\hat A=\hat\Omega(f_A)=\int f_A(x) \hat D(x)  {\mathrm d} x.
\label{quant-map}
\end{gather}
The map $\hat\Omega$ is often called the quantization map, because of its role in quantum mechanics.
In the standard quantization scheme, which associates Hermitian operators to real functions, it is the Weyl map,
whereas its inverse is the Wigner map.
We assume that an appropriate measure ${\mathrm d} x$ exists to make sense of the reconstruction formula and that the
two maps have a~suf\/f\/iciently rich domain to ensure that the maps are invertible for a~dense set in the space of
functions.
In this way we have an invertible map between functions on space and operators on a~Hilbert space.

The symbols form an associative algebra endowed with a~noncommutative (star) product def\/ined in terms of the operator
product
\begin{gather}
f_A\star f_B (x)=\Omega^{-1}\big(\hat A\hat B \big)=\tr\big(\hat A\hat B \hat U (x)\big),
\label{starprodwithtrace}
\end{gather}
where the associativity follows from the associativity of the operator product.

The star product may be expressed in terms of an integral kernel
\begin{gather}
f_A\star f_B (x)= \int f_A(x') f_B (x'')
K(x', x'', x)   {\mathrm d} x' {\mathrm d} x'',
\label{stprod}
\end{gather}
with
\begin{gather}
K(x', x'', x)= \tr\big(\hat D(x') \hat D(x'') \hat U(x)\big).
\label{kernel}
\end{gather}
The operators $\hat D(x)$ and $\hat U(x)$ are also known as quantizer and dequantizer.
The associativity condition implies
\begin{gather*}
\int K(x', y, x) K(x'', x''', y)  {\mathrm d} y = \int K(x',x'', y) K(y, x''', x)   {\mathrm d} y.
\end{gather*}
From the compatibility request of equations~\eqref{deq-map},~\eqref{quant-map} we easily derive an important condition that
the two families of basic operators have to satisfy
\begin{gather}
\tr \big(\hat D(x') \hat U(x)\big) = \delta(x'- x),
\label{compatibility}
\end{gather}
where $\delta$ is to be replaced with the Kronecker delta for discrete parameters.
Going back to~\eqref{kernel}, it is to be stressed that the kernel of the star product has been obtained solely in terms
of the operators~$D(x)$ and~$U(x)$, which in turn are only constrained by~\eqref{compatibility}.
Thus, to each pair of operators satisfying~\eqref{compatibility} it is associated an associative algebra with a~star
product.
The role of $D(x)$ and $U(x)$ can be exchanged, what gives rise to a~duality symmetry~\cite{marmo2}.
This duality allows for the def\/inition of a~new star product, dif\/ferent from the one in~\eqref{stprod} (unless $\hat U$
and $\hat D$ are proportional), def\/ined in terms of a~dual kernel
\begin{gather*}
K^{d}(x',x'',x)=\tr \big(\hat U(x') \hat U(x'') \hat D(x)\big).
\end{gather*}

\section[The Weyl--Wigner and the $s$-ordered maps]{The Weyl--Wigner and the $\boldsymbol{s}$-ordered maps} \label{Section3}

In this section we review the usual Moyal product on the two-dimensional plane ${\mathbb R}^2$ and the family of
$s$-ordered products which generalizes it.
The generalization is made in terms of the quantization-dequantization maps illustrated above.

We f\/irst review a~few properties of coherent states and their relation to the number basis, which shall be used in the
rest of this section.
Let $x=(x_1,x_2)\in {\mathbb R}^2$.
We use a~slightly unconventional notation for the coordinates on the complex plane def\/ining:
\begin{gather*}
z=\frac1{\sqrt{2}}(x_1+{\mathrm i} x_2).
\end{gather*}
We def\/ine the creation and annihilation operators on the plane, $a^\dag$, $a$, as
\begin{gather*}
a^\dag= \frac1{\sqrt 2} (\hat x_1- {\mathrm i}\hat x_2),
\qquad
a = \frac1{\sqrt 2} (\hat x_1+{\mathrm i}\hat x_2)
\end{gather*}
with commutation relation
\begin{gather*}
[a, a^\dag]=\theta,
\end{gather*}
where $\theta$ is a~dimensional parameter of area dimension.
{\it Coherent states} of the plane are thus def\/ined by $a|z\rangle=z|z\rangle$.
The decomposition of the identity reads
\begin{gather*}
\mathbf{1}=\frac{1}{\pi\theta}\int {\mathrm d}^2z |z\rangle\langle z|.
\end{gather*}
Coherent states are non-orthogonal
\begin{gather*}
\langle z|z'\rangle = {\mathrm e}^{-\frac{1}{\theta}(\bar z z +\bar z' z' - 2 \bar z z')}.
\end{gather*}
Given the number operator $\hat N=a^\dag a$, and indicating with $|n\rangle$ its eigenvalues, we state the useful
relations
\begin{gather}
a|n\rangle= \sqrt{n\theta}|n-1\rangle,
\qquad
a^\dag |n\rangle = \sqrt{(n+1)\theta}|n+1\rangle,
\qquad
|n\rangle= \frac{\big(a^\dag\big)^n}{\sqrt{n!\theta^n}} |0\rangle
\label{eigen-n}
\end{gather}
together with
\begin{gather*}
\langle z|n\rangle= {\mathrm e}^{-\frac{\bar z z}{2\theta}} \frac{\bar z^n}{\sqrt {n! \theta^n}}.
\end{gather*}

\subsection{The Moyal product}\label{Section3.1}

Given an operator $\hat A$ acting in the Hilbert space of square integrable functions on ${\mathbb R}$, the Wigner--Weyl
symbol of $\hat A$ and the reconstruction map are def\/ined by means of the families of operators
\begin{gather}
\hat U_{\rm M}(x) = 2~\hat {\mathcal D}(z,\bar z) (- \hat 1)^{a^\dag a} \hat {\mathcal D}(-z, -\bar z),
\label{wigne}
\\
\hat D_{\rm M}(x) = {1\over (2\pi)^2} \hat U_{\rm M}(x),
\label{weyl}
\end{gather}
where $(-\hat 1)^{a^\dag a}$ is the parity operator.
To simplify notations we assume $\theta=1$ unless otherwise stated.
$\hat{\mathcal D} (z,\bar z)$ is the unitary
displacement operator realizing the ray representation of the group of translations of the plane
\begin{gather}
\hat{\mathcal D} (z, \bar z) = {\rm e}^{z a^\dag-\bar z a},
\qquad
\hat{\mathcal D} (z, \bar z) |w\rangle = |w+z\rangle {\rm e}^{\frac{1}{2}(z\bar w- w\bar z)}.
\label{displacement}
\end{gather}

We shall use the isomorphism ${\mathbb R}^2= {\mathbb C}$ in the following and the notation $\hat{\mathcal D} (z)$ as
a~shorthand for $\hat{\mathcal D}(z,\bar z)$ to indicate the functional dependence on the plane coordinates from now on.

The compatibility condition~\eqref{compatibility} between the operators~\eqref{wigne} and~\eqref{weyl} is then readily
verif\/ied on using well known properties of $\hat{\mathcal D} (z)$
\begin{gather}
(-\hat 1)^{a^\dag a} \hat {\mathcal D}(z) (-\hat 1)^{a^\dag a}= \hat {\mathcal D}(-z),
\qquad
\tr\hat {\mathcal D}(z)=\pi \delta({\rm Re}~z) \delta(\operatorname{Im}z).
\label{displtrace}
\end{gather}
The quantizer and dequantizer may be expressed in the more familiar form
\begin{gather}
\hat D_{\rm M}(x)=\frac{1}{(2\pi)^2} \hat U_{\rm M}(x)= \int \frac{{\mathrm d}^2 \xi}{(2\pi)^2} \exp\big[{\mathrm i}\big(\xi_1(\hat x_2-x_2)
+\xi_2(\hat x_1-x_1)\big)\big].
\label{quantizer}
\end{gather}
The Weyl quantization map obtained by the reconstruction formula~\eqref{quant-map} therefore reads
\begin{gather*}
\hat \Omega(f)=\hat f= \int {\mathrm d}^2 x f(x) \int \frac{{\mathrm d}^2 \xi}{(2\pi)^2} \exp\big[{\mathrm i}\big(\xi_1(\hat
x_2-x_2) +\xi_2(\hat x_1-x_1)\big)\big],
\end{gather*}
whereas the inverse map obtained by~\eqref{quant-map} is represented by:
\begin{gather*}
f(x) =
\tr\left(\hat f\int{\mathrm d}^2\xi\exp\big[{\mathrm i}\big(\xi_1(\hat x_2-x_2) +\xi_2(\hat x_1-x_1)\big)\big]\right).
\end{gather*}
The algebra of symbols so def\/ined is endowed with the well known Moyal product.
In this language it is def\/ined in terms of the kernel, which is easily obtained specializing~\eqref{kernel} to this
case.
On restoring the noncommutative parameter~$\theta$ and using the f\/irst of equations~\eqref{displtrace} together with the
product rule for the displacement operators
\begin{gather*}
\hat {\mathcal D} (z)\hat {\mathcal D}(z')=\hat {\mathcal D}(z+z') {\mathrm e}^{{\mathrm i} \operatorname{Im}(z\bar z')}
\end{gather*}
we have indeed
\begin{gather}
K_{\rm M}(x',x'',x)= C \exp\big\{{2{\mathrm i}} \epsilon^{ij} \big[
x_i x'_j+ x'_i x''_j+ x''_i x_j\big] \big\}
\label{moykernel}
\end{gather}
with~$C$ a~normalization constant and $\epsilon^{ij}$ the antisymmetric tensor.
Its expression in complex coordinates is also useful for future comparison
\begin{gather}
K_{\rm M}(z',z'',z)
= C \exp\big\{4{\mathrm i} \operatorname{Im} \big[\bar z'z''+z\bar z''+z'\bar z\big] \big\}.
\label{moycompl}
\end{gather}
The Moyal product is then def\/ined as
\begin{gather}
(f\star_{\rm M} g) (x) = \int{{\mathrm d}^2 x'}{{\mathrm d}^2 x''} f (x') g (x'') K_{\rm M}(x',x'',x).
\label{moyal}
\end{gather}
In complex coordinates the relation~\eqref{moyal} reads
\begin{gather*}
(f\star_{\rm M} g)(z,\bar z)
=\int{\mathrm d}^2 z'{\mathrm d}^2 z''f(z',\bar z') g(z'', \bar z'')
{\mathrm e}^{\{{4{\mathrm i}} \operatorname{Im}  [\bar z'z''+z\bar z''+z'\bar z ]  \}}.
\end{gather*}
In the case of the coordinate functions on $\mathbb R^2$, upon restoring the factors of~$\theta$, this gives the usual
star-commutation relation
\begin{gather}
[x_1,x_2]_{\star_{\rm M}}={\mathrm i}\theta
\label{xcomm}
\end{gather}
or equivalently
\begin{gather}
[z,\bar z]_{\star_{\rm M}}= \theta.
\label{zzbarcomm}
\end{gather}
The Moyal product can be given the popular form
\begin{gather}
(f \star_{\rm M} g)(z,\bar z)=f(z,\bar z) \exp
\left[\frac{\theta}{2}\big(\overleftarrow{\partial_{z}} \overrightarrow{\partial_{\bar z}} -\overleftarrow{\partial_{\bar z}}
\overrightarrow{\partial_{z}}\big)\right] g(z,\bar z),
\label{moy}
\end{gather}
where the operator $\overleftarrow{\partial}$ (resp.\ $\overrightarrow{\partial}$) acts on the left (resp.\ on the
right).
It is well known that this form is obtained by the integral expression through an asymptotic expansion in the
parameter~$\theta$~\cite{EGBV}.

The generalization to the higher-dimensional cases is straightforward.

What is properly def\/ined as the Moyal algebra is $\mathcal{M}_\theta:= \mathcal{M}_L({{\mathbb R}}_\theta^2)\cap
\mathcal{M}_R({{\mathbb R}}_\theta^2)$ where $\mathcal{M}_{L}({{\mathbb R}}_\theta^2)$, the left multiplier algebra, is
def\/ined as the subspace of tempered distributions that give rise to Schwartz functions when left multiplied by Schwartz
functions; the right multiplier algebra~$\mathcal{M}_R({{\mathbb R}}_\theta^2)$ is analogously def\/ined.
For more details we refer to the appendix in~\cite{selene} and re\-fe\-rences therein.
In the present article we can think of $\mathcal{M}_\theta$ either as the algebra of $\star$-polynomial functions in
$z$, $\bar z$ properly completed, or the algebra generated by Schwarz functions, or a~variant.
One should however pay attention since the domains of def\/inition depend on the particular form of the product.
Its commutative limit, $\mathcal{F}({\mathbb R}^2)$, is the commutative multiplier algebra $\mathcal{O}_{\rm M}({\mathbb R}^2)$,
the algebra of smooth functions of polynomial growth on ${\mathbb R}^2$ in all derivatives~\cite{GGISV04}.

\subsection[$s$-ordered symbols]{$\boldsymbol{s}$-ordered symbols}\label{Section3.2}

$s$-ordered symbols have been introduced long ago by Cahill and Glauber~\cite{cahillglauber} in the context of quantum
mechanics to deal with dif\/ferent orderings w.r.t.
the Weyl--Wigner symmetric ordering of operators.
The corresponding quantization and dequantization maps are often referred to as {\it weighted} maps.
We refer to~\cite{MMV04} and references therein for more details.
Here the stress will be on the dif\/ferent star-products which are related to such orderings.
We consider the two families of operators $\hat U_s(x)$ and $\hat D_s(x)$, $x\in {\mathbb R}^2$, of the
form~\cite{cahillglauber}
\begin{gather}
\hat U_s(x) = {2\over (1+s)} \hat {\mathcal D}(z) \left({s-1\over s+1}\right)^{a^\dag a} \hat {\mathcal D}(-z)=
\frac{2}{1+s}\left(\frac{s-1}{s+1}\right)^{(a^\dag-\bar z)(a-z)},
\label{s-deq}
\\
\hat D_s(x) = {1\over (2\pi)^2} \hat U_{-s}(x),
\label{sordop}
\end{gather}
where~$s$ is chosen to be real in the analysis below.
We shall see that the convergence of the kernel will impose constraints on the allowed values of~$s$.
The case $s=0$ corresponds to the standard Wigner--Weyl situation described above.
Two other relevant cases correspond to the singular limits $s=\pm 1$ which yield respectively normal and anti-normal
ordering in quantization.
It is interesting to notice that these two cases are in the duality relation discussed previously (namely the role of
$\hat U$ and $\hat D$ are exchanged).
The duality symmetry then connects normal and anti-normal ordering.
The Wigner--Weyl Moyal quantization scheme selects instead the symmetric ordering, consistently with it being self-dual.

For $s\ne \pm 1$ the star-product kernel of $s$-ordered symbols is calculated along the same lines as the Moyal
kernel~\eqref{moykernel}.
To our knowledge it was f\/irst calculated in~\cite{marmo2}, although no analysis on the convergence was
made.
Let us sketch the derivation.
For simplicity we def\/ine $q= (s+1)/(s-1)$, $\tilde q= q^{-1}$ and we skip overall constants.
From the def\/inition~\eqref{kernel} and the dequantizer and quantizer operators~\eqref{s-deq},~\eqref{sordop}, we have
\begin{gather*}
K_s(z',z'',z) = \tr \big(\hat D (\bar z', z') \hat D (\bar z'', z'') \hat U(\bar z, z) \big)=(1-q)^2 (1-\tilde q)
\\
\qquad
\times \tr \big(\hat {\mathcal{D}} (-\bar z, -z) \hat{\mathcal{D}} (\bar z',z') q^{a^\dag a} \hat{\mathcal{D}} (-\bar z',-z')
\hat{\mathcal{D}} (\bar z'',z'') q^{a^\dag a} \hat{\mathcal{D}}(-\bar z'', -z'') \hat{\mathcal{D}} (\bar z, z){\tilde q}^{a^\dag a} \big).
\end{gather*}
Let us introduce
\begin{gather}
x= -z+z',
\qquad
y=-z'+z''
\qquad
\text{so that}
\qquad
-z''+z=-x-y.
\label{xy}
\end{gather}
On using
the composition rule for the generalized displacement operator $\hat {\mathcal{D}} (\bar w, z)= {\rm{exp}}(z
a^\dag- \bar w a)$
\begin{gather*}
\hat {\mathcal{D}}(\bar w, z) \hat {\mathcal{D}} (\bar u, y)= \hat {\mathcal{D}}(\bar w+\bar u, z+y)
\exp\big(\tfrac{1}{2}(z \bar u- y \bar w) \big)
\end{gather*}
and the commutation relation
\begin{gather*}
q^{a^\dag a} \hat{\mathcal{D}}(\bar z, z)= \hat{\mathcal{D}}(\tilde q\bar z, qz) q^{a^\dag a},
\end{gather*}
we arrive at
\begin{gather}
K_s(z',z'',z) = K_s(x,y)= (1-q) (1-\tilde q)^2 \tr\big[\hat{\mathcal{D}}((1-\tilde q)\bar x, (1-q)x)
q^{a^\dag a}\big]
\nonumber
\\
\hphantom{K_s(z',z'',z) =}{}
\times
\exp
\big[(1-q)\bar x y -(1-\tilde q)\bar y x -\tfrac{1}{2}(q-\tilde q)\bar x x\big].
\label{intermker}
\end{gather}
On using the coherent states basis to compute the trace, we have
\begin{gather*}
q^{a^\dag a}|w\rangle= {\rm e}^{(q^2-1)\frac{\bar w w}{2}} |q w\rangle,
\end{gather*}
so that we are left with a~Gaussian integral
\begin{gather*}
\tr\big[\hat{\mathcal{D}}((1-\tilde q)\bar x, (1-q)x)
q^{a^\dag a} \big]= {\rm e}^{(\tilde q -q) \bar x x}\int {\mathrm d}^2 w
{\rm e}^{-(1-q) (\bar w-\bar x)(w-x)} = \frac{{\rm e}^{(\tilde q -q) \bar x x}}{1-q}.
\end{gather*}
On replacing into equation~\eqref{intermker} we f\/inally have
\begin{gather}
K_s(x,y)= C (1-\tilde q)(1-q) {\rm e}^{(\tilde q -q) \bar x x+ (1-q) \bar x y -(1-\tilde q) \bar y x}
\label{beautiful}
\end{gather}
with~$C$ an overall constant.
This expression can be seen to reproduce the Moyal kernel~\eqref{moycompl} for $s=0$ ($q=\tilde q= -1$).
On replacing the original variables $z$, $z'$, $z''$, it can be seen to coincide with the result of~\cite{marmo2}, up to the
exchange of~$q$ with $\tilde q$.
We have indeed
\begin{gather}
K_s(z',z'',z)= C (1-\tilde q)(1-q) {\rm exp}\big[(\tilde q- q) \bar z z +(\tilde q-1) z'\bar z''+(1-q) z'' \bar
z'+(q-1) z''\bar z
\nonumber\\
\phantom{K_s(z',z'',z)=}{}
+ (1-\tilde q) z\bar z''+ (q-1) z \bar z' + (1-\tilde q) z'\bar z \big].\label{missingref}
\end{gather}
Let us notice that this expression is singular for $s\rightarrow \pm 1$.
These two cases have to be considered separately (in the following we will explicitly compute the kernel for the case
$s=1$). We will denote with $\star_s$ the corresponding star product and with $\mathcal{A}_s$ the noncommutative algebra
of functions on the plane with $\star_s$-noncommutativity, so to have
\begin{gather*}
(f\star_s g)(\bar z, z) = \int {\mathrm d}^2 z'  {\mathrm d}^2 z''
f(\bar z',z') g(\bar z'', z'') K_s(z',z'',z).
\end{gather*}
The convergence of such an expression might impose severe constraints on the algebra of functions selected and has to be
analyzed case by case.
A~careful analysis of the convergence of the kernel shows that its integral is f\/inite and equal to 1 (as it should) only
for $-1<s\le 0$.
Indeed the introduction of the variables~$x$ and~$y$ makes it clear that for such a~choice of~$s$ the integral of
$K(x,y)$ is Gaussian.

As for series expressions of the product, analogues of~\eqref{moy} it should be possible to obtain them on restoring the
noncommutative parameter and expanding in terms of it, as in the Moyal case.
We are not aware that such expressions have been investigated before, but we see no a~priori obstruction to perform the
calculation.

To illustrate the usefulness of equation~\eqref{beautiful} let us compute the star product of~$z$ with $\bar z$ in detail.
We have
\begin{gather*}
z\star_s \bar z= \int {\mathrm d}^2 z'  {\mathrm d}^2 z''
z' \bar z'' K_s(z',z'',z).
\end{gather*}
On expressing $z'$, $z''$ in terms of the variables~$x$ and~$y$ introduced in~\eqref{xy} and replacing the integral kernel
we arrive at
\begin{gather*}
z\star_s \bar z= (1-\tilde q)(1-q) \int {\mathrm d}^2 x  {\mathrm d}^2 y
(x+z)(\bar x+\bar y +\bar z) {\rm e}^{(\tilde q -q) \bar x x+ (1-q) \bar x y -(1-\tilde q) \bar y x},
\end{gather*}
which may be reduced to a~sum of Gaussian integrals.
By direct calculation we see that the integrals containing $\bar x x$, $\bar x z$, $\bar z x$ and $\bar y z$ are zero.
We are left with
\begin{gather}
z\star_s \bar z= (1-\tilde q)(1-q) \int {\mathrm d}^2 x  {\mathrm d}^2 y
(\bar z z + x \bar y) {\rm e}^{(\tilde q -q) \bar x x+ (1-q) \bar x y -(1-\tilde q) \bar y x}.
\label{zbarzprod}
\end{gather}
The f\/irst integral is exactly the integral of the kernel, multiplied by the factor $\bar z z$
\begin{gather}
\int {\mathrm d}^2 x  {\mathrm d}^2 y
{\rm e}^{(\tilde q -q) \bar x x+ (1-q) \bar x y -(1-\tilde q) \bar y x} = \frac{1}{(1-q)(1-\tilde q)}.
\label{part1}
\end{gather}
The second one can be seen to yield
\begin{gather}
\int {\mathrm d}^2 x {\mathrm d}^2 y
\bar y x
{\rm e}^{(\tilde q -q) \bar x x+ (1-q) \bar x y -(1-\tilde q) \bar y x} = \frac{1}{(1-q)(1-\tilde q)^2}.
\label{part2}
\end{gather}
Thus, replacing~\eqref{part1} and~\eqref{part2} into~\eqref{zbarzprod}, we obtain the interesting result
\begin{gather*}
z\star_s \bar z= \bar z z + \frac{1}{1-\tilde q}.
\end{gather*}
Analogously we obtain
\begin{gather*}
\bar z \star_s z= \bar z z - \frac{1}{1- q}
\end{gather*}
so to obtain
for the star commutator
\begin{gather*}
z\star_s \bar z -\bar z \star_s z= 1,
\end{gather*}
which is independent of~$q$ as expected (we recall that all translation-invariant products are equivalent from the point
of view of the star commutator, being a~quantization of the same Poisson bracket).

In this section we have chosen for simplicity to stick to~$s$, hence $q$, $\tilde q$, real.
The analysis can be repeated for~$s$ complex.
The convergence of the kernel and of the star product will impose constraints on the real part of~$s$.

\subsection{The Wick--Voros product} \label{Section3.3}

Let us consider the normal ordered case ($s=1$) in more detail.
It has been shown~\cite{cahillglauber} that in such case the limit for the quantizer is singular (its eigenvalues are
inf\/inite for all values of~$z$) and we may represent it as the normal-ordered operator delta function
\begin{gather}
\hat D_{\rm W} (z)= \int {{\mathrm d}^2 \xi} {\mathrm e}^{{\xi} (a^\dag-\bar z)} {\mathrm e}^{-\bar\xi (a- z)}
\label{quantizervoros}
\end{gather}
with $\xi = \xi_1+{\mathrm i}\xi_2$, which dif\/fers from~\eqref{quantizer} by a~phase.
Operators in such a~scheme acquire the normal-ordered expression
\begin{gather*}
\hat f=\sum\limits_{n,m=0}^\infty f_{nm} \big(a^{\dag}\big)^n a^m.
\label{f_V}
\end{gather*}
The dequantizer instead is simply the projection operator over coherent states
\begin{gather*}
\hat U_{\rm W}(z)= |z\rangle\langle z|,
\end{gather*}
so that the Wick--Voros star product which is associated to this quantization-dequantization scheme, can be easily
computed by means of equation~\eqref{starprodwithtrace} in terms of the expectation value over coherent states
\begin{gather}
(f \star_{\rm W} g)(z,\bar z)=\tr \big(|z\rangle\langle z|\hat f\hat g|
\big)= \langle z|\hat f \hat g|z\rangle =
\sum\limits_{n,m,p,q} f_{nm} g_{pq} \langle z|\big(a^{\dag}\big)^n a^m \big(a^{\dag}\big)^p a^q |z\rangle
\nonumber
\\
\phantom{(f \star_{\rm W} g)(z,\bar z)}
= \int{{\mathrm d}^2 \xi}\sum\limits_{n,m,p,q} f_{nm} g_{pq}\bar z^n z^q \langle z| a^m |\xi\rangle\langle \xi|
\big(a^{\dag}\big)^p |z\rangle
\nonumber
\\
\phantom{(f \star_{\rm W} g)(z,\bar z)}
= \int{{\mathrm d}^2 \xi}\sum\limits_{n,m,p,q} f_{nm} g_{pq}\bar z^n z^q \xi^m \bar\xi^p \langle z
|\xi\rangle\langle\xi| \big(a^{\dag}\big)^p |z\rangle
\nonumber
\\
\phantom{(f \star_{\rm W} g)(z,\bar z)}
= \int {{\mathrm d}^2 \xi} f(\bar z, \xi) g(\bar \xi, z)
|\langle z|\xi\rangle|^2.
\label{vorosprod}
\end{gather}
Alternatively, on using equation~\eqref{kernel}, we may compute the kernel of the star product
\begin{gather*}
K_{\rm W}(z',z'',z) = \tr \big[\hat D(\bar z',z') \hat D(\bar z'',z'') \hat U(\bar z,z)\big]=\langle z| \hat D(\bar z',z')
\hat D(\bar z'',z'')|z \rangle
\nonumber
\\
\phantom{K_{\rm W}(z',z'',z)}{}
 = \int {{\mathrm d}^2 \xi} \langle z | \hat D(\bar z',z')| \xi\rangle \langle \xi | \hat D(\bar z'',z'')| z\rangle
\end{gather*}
and replace the latter in the star product expression
\begin{gather}
(f\star_{\rm W} g)(\bar z,z)= \int {\mathrm d}^2z'{\mathrm d}^2 z'' f(\bar z',z') g(\bar z'', z'') K_{\rm W}(z',z'',z).
\label{vpro}
\end{gather}
On using for example the matrix basis expansion of operators equation~\eqref{f_V} and the quantization map
equation~\eqref{quant-map} we then observe that
\begin{gather*}
f(\bar z, \xi) \langle z|\xi\rangle = \langle z| \hat f |\xi\rangle = \int {\mathrm d}^2 z' f(\bar z',z') \langle z|
\hat D(\bar z',z')|\xi\rangle,
\end{gather*}
which shows that equation~\eqref{vpro}
coincides with the direct computation obtained in~\eqref{vorosprod}.

As in the case of the Moyal product, upon restoring the parameter~$\theta$ we can perform a~series expansion, yielding
the popular expression
\begin{gather}
(f\star_{\rm W} g)(z,\bar z)= f(z,\bar z) \exp \big[\theta
\overleftarrow{\partial_{z}}  \overrightarrow{\partial_{\bar z}} \big] g(z,\bar z).
\label{vorosasympt}
\end{gather}
The Wick--Voros product has been employed in QFT to discuss the emergence of the mixing independently from the specif\/ic
translation invariant product chosen~\cite{Galluccio0}.
In~\cite{Basu} the Wick--Voros star product is singled out as it allows for a~consistent def\/inition of quantum state.

A word of caution is in order, concerning the domain and the range of the weighted Weyl map associated to the Wick--Voros
quantizer~\eqref{quantizervoros}.
While the standard Weyl map associates to Schwarzian functions Hilbert Schmidt operators, for the weighted Weyl map
determined the Wick--Voros quantizer~\eqref{quantizervoros} this is not always the case.
Explicit counterexamples are discussed in~\cite{fuzzydisc2,fuzzydisc1,fuzzydisc3,fuzzydisc4}.
The exact correspondence between the appropriate subalgebras of smooth functions on the plane and bounded operators is
discussed in~\cite{soloviev} for all $s$-ordered quantization schemes, whereas the convergence of the series expansion
in~\eqref{vorosasympt} has been extensively discussed and established in~\cite{waldmann} and references therein.

One important aspect of these products is the fact that derivatives are inner automorphisms:
\begin{gather*}
\frac\partial{\partial x^i} f=\theta^{-1}\varepsilon_{ij}[x^j,f]_\star
\end{gather*}
and
\begin{gather}
\partial_z f=\theta^{-1}[f,\bar z]_\star,
\qquad
\partial_{\bar z} f=\theta^{-1}[f,z]_\star.
\label{derivcommz}
\end{gather}
Note that these relations are valid for all products considered in this section.
This is a~straightforward consequence of the fact that the $\star$ commutator~\eqref{zzbarcomm} holds not only for the
Moyal product, but for all $s$-ordered products.

\subsection{Translation invariance}\label{Section3.4}

Def\/ining the translation in the plane ${\mathbb R}^2$ by a~vector~$a$ as
${\mathcal T}_a(f)(x)=f(x+a)$, by translation invariant product we mean the property
\begin{gather}
{\mathcal T}_a(f)\star {\mathcal T}_a(g)={\mathcal T}_a(f\star g).
\label{tril}
\end{gather}
It is interesting to notice that $s$-ordered star products described in the previous subsection are translation invariant.
This is an almost obvious consequence of the fact that $s$-ordered star products are def\/ined in terms of the displacement
operator~\eqref{displacement},
which realizes a~representation of the group of translations of the plane.
It is however straightforward to check the relation~\eqref{tril}, once we observe that the integral kernel for $s$-ordered
products~\eqref{missingref} verif\/ies
\begin{gather*}
K(z', z'', z+\xi) = K(z'-\xi, z''-\xi, z).
\end{gather*}
A~slightly more general form for translation invariant products of the plane, which also includes commutative ones is
represented~by
\begin{gather*}
(f\star g)(x)=\frac1{2\pi}\int{\mathrm d}^2p  {\mathrm d}^2q
{\mathrm e}^{{\mathrm i} p \cdot x} \tilde f(q)\tilde g(p-q) {\mathrm e}^{\alpha(p,q)}
\end{gather*}
with $\tilde f$, $\tilde g$ the Fourier transforms of $f$, $g$.

The function~$\alpha$ is further restricted by the associativity request.
A~full analysis of the family of translation-invariant products, together with a~study of the cohomology associated to
them, is performed in~\cite{galluccio,varshovi}.
The usual pointwise product is reproduced by $\alpha=0$, the Moyal product by $\alpha_{\rm M} (p,q)=-{\mathrm
i}/2\theta^{ij}q_ip_j$ and the Wick--Voros product by $\alpha_{\rm W}(p,q)=-\theta q_-(p_+-q_+)$, with $q_\pm=\frac{q_1\pm
{\mathrm i} q_2}{\sqrt 2}$.

\section[Matrix bases for $s$-ordered products]{Matrix bases for $\boldsymbol{s}$-ordered products}\label{Section4}

In this section we consider matrix bases for the $s$-ordered star-products described in the
previous sections.
We shall give a~unif\/ied derivation for all of them and then specialize to the known cases of the Moyal~\cite{pepejoe,pepejoe+}
and Wick--Voros~\cite{fuzzydisc1} matrix bases.

To a~function on the plane $\mathbb{R}^2$ we associate via the quantization map~\eqref{quant-map} and the $s$-ordered
quantizer~\eqref{sordop} the $s$-ordered operator
\begin{gather*}
\phi (z,\bar z)\rightarrow:\hat \phi (a, a^\dag):_s
\end{gather*}
with $:\;
:_s$ denoting $s$-ordering.
This may be expanded into $s$-ordered powers of $a$, $a^\dag$
\begin{gather}
\hat\phi =\sum\limits_{p,q} \tilde\phi_{p q}:\big(a^\dag\big)^p a^q:_s.
\label{phi'}
\end{gather}
On using the number basis def\/ined in equation~\eqref{eigen-n} we may rewrite~\eqref{phi'} as
\begin{gather*}
\hat\phi =\sum\limits_{p,q} \phi^{s}_{p q} |p\rangle\langle q|
\end{gather*}
with $\tilde\phi_{pq}$, $\phi^{s}_{k l}$ related by a~change of basis which depends explicitly on the ordering.
On applying the dequantization formula~\eqref{deq-map} with the $s$-ordered dequantizer def\/ined by equation~\eqref{s-deq} we
obtain a~function in the noncommutative algebra $\mathcal{A}_s$
\begin{gather}
\phi(z, \bar z) =\sum\limits_{p,q} \phi^s_{p q} f^s_{pq}(z, \bar z)
\label{basexp}
\end{gather}
with
\begin{gather*}
f^s_{pq}(z, \bar z)= \tr \big(|p\rangle\langle q| \hat U_s(\bar z, z)\big)= \frac{1}{\sqrt{p!q!
\theta^{p+q}}}\tr\big(\big(a^\dag\big)^p|0\rangle\langle 0| a^q \hat U_s(\bar z, z)\big).
\end{gather*}
By def\/inition (cf.~\eqref{starprodwithtrace}) this yields $f^s_{pq}(z, \bar z)$ as a~star product
\begin{gather}
f^s_{pq}(z, \bar z)= \frac{1}{\sqrt{p!q!\theta^{p+q}}}\bar z^p \star_s f_{00}\star_s z^q
\label{wigner functions}
\end{gather}
with $f^s_{00}(\bar z, z)$ the $s$-ordered symbol of the operator $|0\rangle\langle 0|$
\begin{gather*}
f^s_{00}(\bar z, z)= \tr\big(|0\rangle\langle 0| \hat U_s(\bar z,z)\big)
\end{gather*}
and $\bar z^p$, $z^q$ respectively symbols of $\big(a^\dag\big)^p$, $a^q$ in all schemes (equation~\eqref{wigner functions} is thus true
by def\/inition of star product~\eqref{starprodwithtrace}, and the use of associativity:
\begin{gather*}
(f_{A}\star_s f_B\star_s f_C)(\bar z,z)=\tr \big(\hat A\hat B\hat C \hat U_s(\bar z,z)\big)
\end{gather*}
with the identif\/ication $\hat A= \big(a^\dag\big)^p$, $\hat B= |0\rangle\langle 0|$, $\hat C= a^q$).

It is immediate to verify that $f^s_{00}(\bar z, z)$ is idempotent independently from the particular form of the
operator $\hat U_s$. We have indeed
\begin{gather*}
(f_{00}\star_s f_{00}) (\bar z, z) = \tr\big(|0\rangle\langle 0 |0\rangle\langle 0| \hat U_s(\bar z,z)\big)=f_{00}(\bar z, z).
\end{gather*}
The basis elements $f_{pq}(z,\bar z)$ may be seen to obey the following fusion rule
\begin{gather}
f_{pq}\star_s f_{kl}=\delta_{qk}f_{pl}
\label{fusion}
\end{gather}
by observing that, by def\/inition
\begin{gather*}
f_{pq}\star_s f_{kl}=\tr\big(|p\rangle\langle q|k\rangle\langle l | \hat U_s(\bar z,z)\big).
\end{gather*}
This implies that every $s$-ordered star product may be described as matrix product.
We have indeed
\begin{gather*}
\phi \star_s\psi (\bar z,z)= \sum\phi_{nm}\psi_{pq} f_{nm} \star_s f_{pq} (\bar z, z)= \sum \phi_{nm}\psi_{mq} f_{nq}
(\bar z, z)= \sum (\Phi \cdot\Psi)_{nq} f_{nq}(\bar z, z)
\end{gather*}
with $\Phi$, $\Psi$ the inf\/inite matrices with entries the series expansion coef\/f\/icients of the func\-tions~$\phi$,~$\psi$
(see equation~\eqref{basexp}).
The idempotent function $f_{00}$ may be computed explicitly.
On using the number basis~$n$ and the $s$-ordered dequantizer~\eqref{s-deq} we have
\begin{gather}
 \frac{(1+s)}{2}f_{00}(\bar z, z)= \sum\limits_n\langle n |0\rangle\langle 0| \hat {\mathcal D}(z) \left({s-1\over
s+1}\right)^{a^\dag a} \hat {\mathcal D}(-z) |n\rangle
\nonumber
\\
\hphantom{\frac{(1+s)}{2}f_{00}(\bar z, z)}{}
= \langle 0| \hat {\mathcal D}(z) \left({s-1\over s+1}\right)^{a^\dag a} \hat {\mathcal D}(-z) |0\rangle =
\sum\limits_{n m} \langle -z| n\rangle \langle n | \left({s-1\over s+1}\right)^{a^\dag a} |m\rangle\langle m| |-z\rangle
\nonumber
\\
\hphantom{\frac{(1+s)}{2}f_{00}(\bar z, z)}{}
= {\mathrm e}^{-\frac{\bar z z}{\theta}} \sum\limits_{nm}(-1)^{n+m} \frac{\bar z^n z^m}{\sqrt {n! m! \theta^{n+m}}}
\left({s-1\over s+1}\right)^{m}\delta_{nm}= {\mathrm e}^{-\frac{\bar z z}{\theta}} {\mathrm
e}^{\frac{s-1}{s+1}\frac{\bar z z}{\theta}}.
\label{idemp}
\end{gather}
As we can see, it depends explicitly on the value of~$s$.
This is a~particular case of a~more general formula~\cite[equations~(6.35),~(6.36)]{cahillglauber}.
We shall see in next sections that it reproduces correctly previous results which have been obtained in specif\/ic
quantization-dequantization schemes.

We can establish the useful result for the integral of the basis functions $f_{pq}$.
We have
\begin{gather}
\int {\mathrm d}^2 z f_{pq} (\bar z, z)=\int{\mathrm d}^2 z \langle q| U_s|p\rangle= 2 \pi\theta \delta_{pq}.
\label{intfpq}
\end{gather}
The generalization to four dimensions is readily obtained on introducing
\begin{gather*}
f_{PQ}(\bar z_a, z_a)=f_{p_1 q_1}(\bar z_1,z_1) \cdot f_{q_2 p_2}(\bar z_2, z_2),
\qquad
a=1,2.
\end{gather*}
Together with the fusion rule equation~\eqref{fusion}, equation~\eqref{intfpq} ensures that the matrix basis $f_{pq} (\bar z, z)$
is orthogonal.
This has the important consequence that the action of every f\/ield theory model with $s$-ordered star product becomes
a~matrix action, with integrals replaced by traces.
To illustrate this point, let us consider for simplicity a~scalar action with polynomial interaction, in two dimensions
\begin{gather*}
S[\phi]= \int {\mathrm d}^2 z (\phi \star_s \widehat O \phi)(\bar z, z) + \lambda \phi^{\star_s n}(z,\bar z)
\end{gather*}
with $\phi^{\star_s n}=\phi\star_s\phi\star_s\cdots \star_s \phi$ n times.
On expanding the f\/ields in the matrix basis as in in equation~\eqref{basexp} we f\/irst observe that, thanks to~\eqref{fusion}
repeatedly applied, the star product in the algebra becomes an inf\/inite-matrix product
\begin{gather*}
\phi_1\star_s \phi_2\star_s \cdots \star_s\phi_n(\bar z, z) = \sum\limits_{p_i, q_i}{\phi_1}_{p_1 q_1} {\phi_2}_{p_2
q_2}\cdots {\phi_n}{p_n q_n} f_{p_1 q_1} \star_s f_{p_2 q_2}\star_s\cdots \star_s f_{p_n q_n}
\\
\qquad
 = {\sum\limits_{p_1, q_n} (\Phi_1\cdot \Phi_2 \cdots  \Phi_n)_{p_1 q_n} f_{p_1 q_n}},
\end{gather*}
where $\Phi=\{\phi_{pq}\}$ are the inf\/inite matrices of f\/ields coef\/f\/icients in the matrix basis
expansion~\eqref{basexp}.
On integrating the latter expression by means of~\eqref{intfpq} we f\/inally get
\begin{gather*}
\int{\mathrm d}^2 z   {\phi_1\star \phi_2}\star\cdots \star\phi_n(\bar z, z)= 2 \pi\theta\tr (\Phi_1\cdot
\Phi_2 \cdots  \Phi_n).
\end{gather*}
For the kinetic term we proceed analogously, although it is in general not diagonal in the matrix basis.
Repeating the same steps we arrive at
\begin{gather*}
\int {\mathrm d}^2 z   (\phi \star_s \widehat O \phi)(\bar z, z) = \tr\Phi O\Phi
\end{gather*}
with
\begin{gather*}
(O)_{pq;rs}=\int {\mathrm d}^2 z  f_{pq}\star_s \widehat O f_{rs}
\end{gather*}
the representation of the kinetic operator on the matrix basis, to be computed case by case.
Applications of this procedure may be found in Section~\ref{Section7}.

\subsection{The Moyal matrix basis} \label{Section4.1}

We have seen in previous sections that the Moyal product is introduced through a~symmetric-ordered quantization scheme.
When qualifying ${\mathbb R}^2$ as the phase space of 1-dimensional systems, the basis functions $f_{pq}(z, \bar z)$ in
this quantization scheme correspond exactly to the Wigner functions associated to the density operator of the quantum
oscillator states.

The Moyal matrix basis has been established long ago by J.M.~Gracia-Bond\'\i a~and J.C.~V\'arilly following a~slightly
dif\/ferent approach~\cite{pepejoe,pepejoe+} with respect to the one described in previous section.
The idempotent function $f_{00}(\bar z, z)$ has been shown~\cite{pepejoe,pepejoe+} to be the Gaussian
\begin{gather*}
f^{\rm M}_{00}(\bar z, z)= 2 \exp(-2\bar z z/\theta),
\end{gather*}
which agrees with our result~\eqref{idemp} at $s=0$.
The expression of the matrix elements $\phi_{k l}$ in terms of $\tilde\phi_{pq}$ has been computed for the Moyal case
in~\cite{LizziSzaboZampini}.

The extension to $\mathbb{R}^4_\theta$ is straightforward.
We have
\begin{gather}
\phi(z_a,\bar z_a)= \sum\limits_{PQ} \phi_{PQ} f^{\rm M}_{PQ}(z_a,\bar z_a)
\label{phir4}
\end{gather}
with $a=1,2$, $P= (p_1,p_2)$ and
\begin{gather*}
f^{\rm M}_{PQ}(z_a,\bar z_a)=f^{\rm M}_{p_1,q_1} (z_1,\bar z_1)\cdot f^{\rm M}_{p_2,q_2} (z_2,\bar z_2).
\end{gather*}
In order to describe elements of $\mathbb{R}_\theta^2$ (resp.\
$\mathbb{R}_\theta^4$), the sequences $\{\phi_{pq}\}$ (resp.\
$\{\phi_{\vec p\vec q}\}$) have to be of rapid decay~\cite{pepejoe,pepejoe+}.

\subsection{The Wick--Voros matrix basis} \label{Section4.2}

We have seen previously that the Wick--Voros product is introduced through a~weighted quantization map which, in two
dimensions, associates to functions on the complex plane normal ordered operators.
The inverse map which is the analogue of the Wigner map is represented by
\begin{gather}
\phi(z,\bar z) =\langle z|\hat \phi|z\rangle.
\label{wigner}
\end{gather}
The Wick--Voros product,
$\phi\star_{\rm W} \psi$, is particularly simple with respect to the other $s$-ordered products
(including the well studied Moyal one).
It is def\/ined as the expectation value over coherent states of the operator product $\hat \phi \hat \psi$.
Then, for analytic functions, a~very convenient way to reformulate the quantization map~\eqref{quant-map} is to consider
the analytic expansion
\begin{gather*}
\phi(\bar z,z)=\sum\limits_ {pq}\tilde\phi_{pq} \bar z^p z^q,
\qquad
p,q\in \mathbb{N},
\end{gather*}
with $\tilde\phi_{pq}\in {\mathbb C}$.
The Wick--Voros quantizer~\eqref{quantizervoros} will then produce the normal ordered operator
\begin{gather*}
\hat \phi=\sum\limits_ {pq}\tilde\phi_{pq} \big(a^{\dag}\big)^p a^q.
\end{gather*}
We will therefore assume analyticity in what follows.
The idempotent function $f_{00}$ is a~Gaussian, as for the Moyal case, although with a~slightly dif\/ferent shape
\begin{gather*}
f^{\rm W}_{00}(\bar z, z)= \exp(-\bar z z/\theta).
\end{gather*}
This result agrees with the general result~\eqref{idemp} for $s=1$.
The basis functions $f^{\rm W}_{pq}$ acquire the simple form
\begin{gather*}
f^{\rm W}_{pq}(z,\bar z)= \frac{{\rm e}^{-\frac{\bar z z}{\theta}}}{\sqrt{p!q!\theta^{p+q}}} \bar z^{p} z^{q},
\end{gather*}
where we notice that no star product is present anymore dif\/ferently from what happens in all other situations described
by equation~\eqref{wigner functions} with $s\ne 1$, including the Moyal case, $s=0$.
This is due to the fact that $\bar z^p \star_{\rm W} f = \bar z^p\cdot f$ as well as $f \star_{\rm W} z^q= f \cdot z^q$.
The generalization to ${\mathbb R}^4$ is straightforward and follows the same lines as for the Moyal case.
We have
\begin{gather}
f^{\rm W}_{PQ}(z_a,\bar z_a)=f^{\rm W}_{p_1,q_1} (z_1,\bar z_1)\cdot f^{\rm W}_{p_2,q_2} (z_2,\bar z_2).
\label{fpq4W}
\end{gather}

\section{Star products as reductions}\label{starprodred} \label{Section5}

The class of products which we have considered up to now is translation invariant, with noncommutative parameters being
constant.
It is interesting to notice that, when considered in four dimensions, through a~reduction procedure these products give
rise to a~whole family of star products in three dimensions, with linear noncommutativity in space coordinates.
This result was f\/irst achieved~\cite{selene} by considering reductions of the Moyal product, while in~\cite{hammou}
a~particular rotation-invariant star product in three dimensions was obtained as a~reduction of the Wick--Voros product.
It turns out that a~reduction in terms of the Wick--Voros product is technically easier to perform, although being
conceptually equivalent.
We will therefore present the reduction in such form.

The crucial step to obtain star products on $\mathcal{F}(\mathbb{R}^3)$, hence to deform $\mathcal{F}(\mathbb{R}^3)$
into a~noncommutative algebra, is to identify $\mathbb{R}^3$ with the dual, $\mathfrak{g}^*$, of some chosen three-dimensional Lie algebra~$\mathfrak{g}$.
This identif\/ication induces on $\mathcal{F}(\mathbb{R}^3)$ the Kirillov--Poisson bracket, which, for coordinate functions reads
\begin{gather}
\{x_i,x_j\}=c_{ij}^k x_k +b_{ij}
\label{Kirillov}
\end{gather}
with $i=1,\dots,3$
and $c_{ij}^k$, $b_{ij}$, the structure constants of~$\mathfrak{g}$.
On the other hand, all three-dimensional (Poisson) Lie algebras may be realized as subalgebras of the inhomogeneous
symplectic algebra $\mathfrak{isp}(4)$, which is classically realized as the Poisson algebra of quadratic-linear
functions on $\mathbb{R}^4$ ($\mathbb{C}^2$ with our choices) with canonical Poisson bracket
\begin{gather*}
\big\{z^a, \bar z^b\big\}= {\mathrm i},
\qquad
a,b=1,2.
\end{gather*}
It is then possible to f\/ind quadratic-linear functions
\begin{gather*}
x_i= x_i(z^a,\bar z^a),
\end{gather*}
which obey~\eqref{Kirillov}.
This is nothing but the classical counterpart of the Jordan--Schwinger map realization of Lie algebra generators in terms
of creation and annihilation operators~\cite{MMVZ}.
Then one can show~\cite{selene} that these Poisson subalgebras are also Wick--Voros (and Moyal) subalgebras, that is
\begin{gather}
x_i(z^a,\bar z^a)\star_{\rm W} x_j(z^a,\bar z^a) -x_j(z^a,\bar z^a)\star_{\rm W} x_i(z^a,\bar z^a)= \lambda \big(c_{ij}^k
x_k(z^a,\bar z^a)+b_{ij}\big),
\label{starx}
\end{gather}
where the noncommutative parameter~$\lambda$ depends on~$\theta$ and shall be adjusted according to the physical
dimension of the coordinate functions $x_i$.
Occasionally we shall indicate with $\mathbb{R}^3_\lambda$ the noncommutative algebra $(\mathcal{F}({\mathbb R}^3),
\star)$.
Equation~\eqref{starx} induces a~star product on polynomial functions on $\mathbb{R}^3$ generated by the coordinate functions
$x_i$, which may be expressed in closed form in terms of dif\/ferential operators on $\mathbb{R}^3$.
For details we refer to~\cite{selene} where all products are classif\/ied.
Here we will consider quadratic realizations of the kind
\begin{gather}
\pi^*(x_\mu)=\kappa \bar z^a e_\mu^{ab} z^b,
\qquad
\mu=0,\dots,3,
\label{xmu}
\end{gather}
with $e_i= \frac{1}{2}\sigma_i$, $i=1,\dots,3$
are the ${\rm SU}(2)$ generators and $\sigma_i$ are the Pauli matrices, while $e_0=\frac{1}{2} \mathbf{1}$.
Here we have explicitly indicated the pull-back map $\pi^*:\mathcal{F}({\mathbb R}^3)\mapsto \mathcal{F}({\mathbb
R}^4)$.
We will shall omit it in the following, unless necessary.
$\kappa$ is some possibly dimensional constant such that $\lambda=\kappa \theta$.
Notice that
\begin{gather*}
x_0^2= \sum\limits_i x_i^2.
\end{gather*}
It is possible to show that the Wick--Voros product on ${\mathbb R}^4$ determines the following star product for the
algebra of functions on ${\mathbb R}^3$, once the ${\rm SU}(2)$ generators have been chosen~\cite{hammou}
\begin{gather}
(\phi \star \psi) (x) = \exp \left[\frac{\lambda}{2}\big(\delta^{ij} x_0 + {\mathrm i} \epsilon_{ij}^k x_k\big)
\frac{\partial}{\partial u_i} \frac{\partial}{\partial v_j}\right] \phi(u) \psi(v) |_{u=v=x}.
\label{starsu2}
\end{gather}
This star product implies for coordinate functions
\begin{gather*}
x_i\star x_j = x_i \cdot x_j + \frac{\lambda}{2}\big(x_0\delta_{ij}+ {\mathrm i}\epsilon_{ij}^kx_k\big),
\qquad
x_0\star x_i=x_i\star x_0= x_0 x_i +\frac{\lambda}{2} x_i,
\\
x_0\star x_0 = x_0 \left(x_0 +\frac{\lambda}{2}\right)= \sum\limits_i x_i\star x_i -\lambda x_0,
\end{gather*}
from which one obtains
\begin{gather*}
[x_i, x_j]_\star = {\mathrm i}{\lambda}\epsilon_{ij}^kx_k.
\end{gather*}
Let us notice that $x_0$ star-commutes with all elements of the algebra, so that it is possible to def\/ine ${\mathbb
R}^3_\lambda$ as the star-commutant of $x_0$.

It is possible to reduce the noncommutative algebra on ${\mathbb R}^4_\theta$ on using dif\/ferent three-dimensional Lie
algebras in equation~\eqref{xmu} or realizations which are not even polynomial~\cite{selene, MMVZ}.
These will give dif\/ferent star products on ${\mathbb R}^3$ which are in general non-equivalent.
In the following we will just consider the star product~\eqref{starsu2} and refer to the corresponding noncommutative
algebra as~${\mathbb R}^3_\lambda$.
The expression~\eqref{starsu2} for the star product in ${\mathbb R}^3_\lambda$ is practically dif\/f\/icult to use in
calculations, for example in QFT.
In next section we shall review a~matrix basis for ${\mathbb R}^3_\lambda$ which makes it much easier to compute the
$\star$-product as it will reduce the $\star$ product~\eqref{starsu2} to matrix multiplication.

\section[Matrix basis for $\mathbb{R}^3_\lambda$]{Matrix basis for $\boldsymbol{\mathbb{R}^3_\lambda}$}\label{R3lambda}\label{Section6}

We review a~matrix basis of $\mathbb{R}^3_\lambda$ which is based on a~suitable reduction of
the matrix basis $f_{PQ}$ discussed in the previous section.

It is well known in the Jordan--Schwinger realization of the ${\rm SU}(2)$ generators, that the eigenvalues of the number
operators $\hat N_1=a^\dag_1 a_1$, $\hat N_2=a^\dag_2 a_2$, say $p_1$, $p_2$, are related to the eigenvalues of
$\hat{\mathbf{X}}^2$, $\hat X_3$, respectively $j(j+1)$ and~$m$,~by
\begin{gather*}
p_1+p_2=2j,
\qquad
p_1-p_2=2m
\end{gather*}
with $p_i\in \mathbb{N}$, $j\in \mathbb{N}/2$, $-j\le m \le j$, so to have
\begin{gather*}
|p_1 p_2\rangle=|j+m, j-m\rangle= \frac{\big(a_1^\dag\big)^{j+m}(a_2)^{j-m}}{\sqrt{(j+m)!(j-m)!}} |00\rangle,
\end{gather*}
where $\hat X_i$, $i=1,\dots,3$
are selfadjoint operators representing the $\mathfrak{su}(2)$ Lie algebra generators on the
Hilbert space spanned by $|j,m\rangle$.
Then we may relabel the matrix basis of $\mathbb{R}^4_\theta$, equation~\eqref{fpq4W} as $f^{j{\tilde j}}_{m \tilde m}$, so to
have
\begin{gather*}
\phi(z_a,\bar z_a)= \sum\limits_{j {\tilde j}\in \mathbb{N}/2}\sum\limits_{m=-j}^j \sum\limits_{\tilde
m=-\tilde{j}}^{\tilde{j}} \phi^{j {\tilde j}}_{m\tilde m} f^{j{\tilde j}}_{m\tilde m}(z_a,\bar z_a).
\end{gather*}
We further observe that, for~$\phi$ to be in the subalgebra $\mathbb{R}^3_\lambda$ we must impose $j=\tilde{j}$.
To this it suf\/f\/ices to compute
\begin{gather*}
x_0 \star f^{j\tilde{j}}_{m\tilde m}-f^{j\tilde{j}}_{m\tilde m}\star x_0=\lambda(j-\tilde{j}) f^{j\tilde{j}}_{m\tilde m}
\end{gather*}
and remember that $\mathbb{R}_\lambda^3$ may be alternatively def\/ined as the $\star$-commutant of $x_0$.
This requires
\begin{gather*}
j=\tilde{j}.
\end{gather*}
We have then
\begin{gather*}
\phi(x_i)=\sum\limits_{j}\sum\limits_{m,\tilde m=-j}^j \phi^j_{m\tilde m} v^j_{m\tilde m}
\end{gather*}
with
\begin{gather*}
v^j_{m\tilde m}:= f^{jj}_{m\tilde m}=\frac{\bar z_1^{j+m} f_{00}(\bar z_1,z_1) z_1^{j+\tilde m} \bar z_2^{j-m}
f_{00}(\bar z_2,z_2) z_2^{j-\tilde m}}{\sqrt{(j+m)!(j-m)! (j+\tilde m)!(j-\tilde m)! \theta^{4j}}}.
\end{gather*}
The orthogonality property now reads
\begin{gather*}
v^j_{m\tilde m}\star v^{\tilde{j}}_{n \tilde n}=\delta^{j \tilde{j}}\delta_{\tilde m n}v^j_{m \tilde n}.
\end{gather*}
As for the normalization we have
\begin{gather*}
\int {\mathrm d}^2 z_1{\mathrm d}^2 z_2
v^j_{m \tilde m} (z, \bar z)= 4 \pi^2 \theta^2 \delta_{m \tilde m}.
\end{gather*}
The star product in $\mathbb{R}_\lambda^3$ becomes a~matrix product
\begin{gather*}
\phi\star \psi (x)=\sum \phi^{j_1}_{m_1\tilde m_1} \phi^{j_2}_{m_2\tilde m_2} v^{j_1}_{m_1\tilde m_1} \star
v^{j_2}_{m_2\tilde m_2} =\sum \phi^{j_1}_{m_1\tilde m_1} \phi^{j_2}_{m_2\tilde m_2} v^{j_1}_{m_1\tilde m_2} \delta^{j^1
j^2} \delta_{\tilde m_1 m_2}
\end{gather*}
while the integral may be def\/ined through the pullback to ${\mathbb R}^4_\theta$
\begin{gather*}
\int_{\mathbb{R}^3_\lambda}{{\mathrm d}^3 x} \phi \star \psi:=\kappa^2 \int_{\mathbb{R}^4_\theta} {{\mathrm d}^4 x}
\pi^\star(\phi)\star_{\rm M} \pi^*(\psi)= 4\pi^2 \lambda^2 \tr \Phi\Psi
\end{gather*}
hence becoming a~trace.

In analogy with the present derivation, the matrix basis adapted to the Moyal product~\cite{pepejoe,pepejoe+} has been reduced to
three dimensions in~\cite{duflo} where applications to quantum mechanics (the hydrogen atom) are considered.

\section{Field theories on noncommutative spaces as matrix models} \label{Section7}

Field theories on noncommutative spaces based on the Moyal product were introduced in~\cite{MinwallavanRaamsdongSeiberg, SeibergWitten}.
It was soon realized that {they could be very ef\/fectively described by matrix models}.
For example it was shown in~\cite{AMNS} that def\/ining a~noncommutative f\/ield theory on a~noncommutative torus (which we
discuss below in Section~\ref{Section8.1}), the theory is def\/ined on a~lattice and becomes a~matrix model of the IKKT
type~\cite{IKKT}.
Another application of matrix bases shows that the gauge Lie algebra of a~theory def\/ined with the Moyal product is
a~particular form of ${\rm SU}(\infty)$~\cite{LizziSzaboZampini} related to the inner automorphisms of the underlying deformed
algebra of functions on spacetime.

In several relevant examples of QFT on noncommutative spaces the introduction of an orthogonal matrix
basis has made it possible to explicitly compute the propagator and the vertices of the models investigated.
We describe some of them in this section.
The importance of the matrix basis is that a~perturbative analysis becomes possible, reducing the problem of loop
calculations to taking traces, which, once regularized with the introduction of a~cutof\/f (so to have f\/inite matrices).
These may be implemented with a~computer program, and some steps in this direction have been taken
in~\cite{SpissoWulkenhaar}.

\subsection{The Grosse--Wulkenhaar model} \label{Section7.1}

An important application of the matrix basis for the Moyal plane is the perturbative analysis of the Grosse--Wulkenhaar
harmonic model~\cite{GrosseWulken1,GrosseWulken2}.
For simplicity we shall only review here the two-dimensional case~\cite{GrosseWulken1} to illustrate the procedure.

The model deals with a~scalar theory with quartic interaction.
It is described by the action
\begin{gather}
S=\int {\mathrm d}^2 z \left(\partial_z \phi \star \partial_{\bar z} \phi + 2\frac{\Omega^2} {\theta^2} \left(z \phi
\star_{\rm M} \bar z \phi+ \bar z \phi \star_{\rm M} z \phi\right) + \frac{1}{2}\mu_0^2 \phi\star_{\rm M} \phi
+\frac{\lambda}{4!}\phi^{\star_{\rm M} 4}\right).
\label{r2action}
\end{gather}
The harmonic term is crucial in four dimensions to cure the famous UV/IR mixing~\cite{MinwallavanRaamsdongSeiberg},
which is quadratic in $d=4$.
It is however worth noting that it breaks the translation invariance of the action.

On using the series expansion~\eqref{phir4} for the f\/ields and the orthogonality properties of the matrix basis
described in Section~\ref{Section4.1}, together with equations~\eqref{derivcommz}, we may rewrite the
action~\eqref{r2action}~as
\begin{gather*}
S=S_{\rm kin}+ S_{\rm int}
\end{gather*}
with
\begin{gather}
S_{\rm int} = \frac{\lambda}{4!}\pi \theta\tr \Phi\cdot \Phi\cdot \Phi\cdot \Phi,
\qquad
S_{\rm kin} = \tr \Phi \Delta \Phi.\label{int}
\end{gather}
Hence we observe that, while the interaction term is polynomial in the matrix $\Phi \equiv (\phi_{mn})$, the kinetic
term is highly non-local (non-diagonal), with
\begin{gather}
\Delta_{mn, kl}= \left(\mu_0^2+2\frac{(1+\Omega^2)}{\theta}(m+n+1)\right)\delta_{nk}\delta_{ml}
-2\frac{(1-\Omega^2)}{\theta}\sqrt{(n+1)(m+1)}\delta_{n+1,k}\delta_{m+1,l}
\nonumber
\\
\phantom{\Delta_{mn, kl}=}
{} - 2\frac{(1-\Omega^2)}{\theta}\sqrt{nm}\delta_{n-1,k}\delta_{m-1,l}.
\label{kinetic}
\end{gather}
The propagator denoted by $P_{mn;kl}$ is the inverse of the kinetic term.
It is def\/ined~by
\begin{gather}
\sum\limits_{k,l}\Delta_{mn;kl}P_{lk;sr}=\delta_{mr}\delta_{ns},
\qquad
\sum\limits_{k,l}P_{nm;lk}\Delta_{kl;rs}=\delta_{mr}\delta_{ns}.
\label{definvers}
\end{gather}
$\Delta$ satisf\/ies an index conservation law
\begin{gather*}
\Delta_{mn;kl}\ne0\iff m+n=k+l.
\end{gather*}
This implies that equation~\eqref{kinetic} depends only on three indices.
Therefore, setting $n=\alpha-m$, $k=\alpha-l$, with $\alpha=m+n=k+l$ we set
\begin{gather}
\Delta_{m,\alpha-m;\alpha-l,l}:=\Delta^{(\alpha)}_{m,l}.
\label{tridiagon}
\end{gather}
One observes that, for each value of~$\alpha$, $\Delta^{(\alpha)}_{ml}$ is an inf\/inite real symmetric tridiagonal matrix
which can be related to a~Jacobi operator.
Therefore, the diagonalization of~\eqref{tridiagon} can be achieved by using a~suitable family of Jacobi orthogonal
polynomials.
This is a~general feature of all subsequent models which shall be analyzed in this section.

Denoting generically by $\lambda_k$, $k\in\mathbb{N}$ the eigenvalues of $\Delta^{(\alpha)}_{mn}$~\eqref{tridiagon}, we
write it as
\begin{gather*}
\Delta^{(\alpha)}_{ml}=\sum\limits_{p\in\mathbb{N}}
{{\mathcal{R}}}^{(\alpha)}_{mp}\left(\frac{2(1+\Omega^2)}{\theta}\lambda_p+\mu_0^2\right){{{\mathcal{R}}}^{(\alpha)}}^\dag_{pl}
\end{gather*}
with
\begin{gather}
\sum\limits_{p\in\mathbb{N}}{{\mathcal{R}}}^{(\alpha)}_{mp}{{{\mathcal{R}}}^{(\alpha)}}^\dag_{pl}=
\sum\limits_{p\in\mathbb{N}}{{{\mathcal{R}}}^{(\alpha)}}^\dag_{mp}{{{\mathcal{R}}}^{(\alpha)}}_{pl}=\delta_{ml},
\label{orthogpoly}
\end{gather}
where ${{{\mathcal{R}}}^{(\alpha)}}^\dag_{mn}={{\mathcal{R}}}^{(\alpha)}_{nm}$.
Then, combining with equation~\eqref{kinetic} we obtain the following 3-term recurrence relation
\begin{gather*}
\big(1-\Omega^2\big) \sqrt{(m+1)(\alpha+m+1)} {{\mathcal{R}}}^{(\alpha)}_{m+1}(\lambda) +\big(1-\Omega^2\big)\sqrt{m(\alpha+m)}{{\mathcal{R}}}^{(\alpha)}_{m-1}(\lambda)
\nonumber
\\
\qquad
{}+\big(\lambda -\big(1+\Omega^2\big)(\alpha+1+2m)\big){{\mathcal{R}}}^{(\alpha)}_{m}(\lambda) =0,
\qquad
\forall\, m,q\in\mathbb{N},
\end{gather*}
where we have traded the discrete index~$q$
for~$\lambda$.
On introducing a~cutof\/f~$N$
on the matrix indices, it has been shown in~\cite{GrosseWulken1} that this is the recurrence
equation for modif\/ied Laguerre polynomials~\cite{kks} $L_m^{\alpha,\omega}(\lambda)$ with
$\omega^{1/2}=(1-\Omega^2)/(1+\Omega^2)$.
The eigenvalues of the Laplacian are the zeroes of the modif\/ied Laguerre polynomials.
The eigenfunctions of the Laplacian are therefore proportional to modif\/ied Laguerre polynomials, up to a~normalization
function, $f(N,\alpha,m)$, which is determined from the orthonormality request.

Once we have diagonalized the kinetic operator, the propagator is readily obtained.
From equation~\eqref{definvers} we obtain
\begin{gather*}
P^{(N,\alpha,\omega)}_{mn}=\sum\limits_{p=0}^{N}f^2(N,m,\alpha)L_m^{\alpha,\omega}(\lambda_p)
\frac{1}{2\frac{(1+\Omega^2)}{\theta}\lambda+\mu_0^2}L_n^{\alpha,\omega}(\lambda_p).
\end{gather*}
The limit $N\to\infty$ is easy to perform in the case $\omega=1$ where the product of Laguerre polynomials gives rise
to the integration measure (see~\cite{GrosseWulken1} for details).

In~\cite{GrosseWulken2} the whole analysis has been repeated for the Moyal space ${\mathbb R}^4_\theta$.
The kinetic term is of the same kind as the one considered here, although in higher dimensions.
It turns out that the recurrence relation which is relevant there, is satisf\/ied by another family of orthogonal
polynomials, the so called Meixner polynomials~\cite{kks}.

\subsection{The translation invariant model}\label{Section7.2}

The translation invariant model has been introduced in~\cite{rivass}.
Its importance resides in the fact that it is renormalizable, while preserving translation invariance.
Indeed, as already noticed in the previous sections, the Moyal star product is an instance of a~translation invariant
product, according to~\eqref{tril}.
This implies that every commutative translation invariant theory keeps such an invariance when deformed by the sole
replacement of the commutative star-product with the Moyal star product or any other translation-invariant one.
However we have seen in the previous section that the noncommutative $\lambda \phi^4$ f\/ield theory is not
renormalizable, unless a~translation invariance breaking term is introduced.
On the other hand, the model brief\/ly described below has the advantage of modifying the propagator of the $\lambda
\phi^4$ model, without destroying the symmetries of its commutative analogue.

The action which describes the model in four dimensions (Euclidean) is
\begin{gather*}
S=\int {\mathrm d}^4 x \left[\frac{1}{2}\left(\partial_\mu\phi \star \partial_\mu \phi + \frac{a}{\theta^2}\partial_\mu^{-1}
\phi \star \partial_\mu^{-1} \phi +m^2 \phi \star \phi\right)\right] + \frac{\lambda}{4!}\phi^{\star 4}
\end{gather*}
with
\begin{gather*}
\partial^{-1}_\mu \phi (x) = \int {\mathrm d} x^\mu \phi(x)= \int {\mathrm d} p \frac 1{{\mathrm i} p^\mu}
\tilde\phi(p){\mathrm e}^{{\mathrm i} p\cdot x}
\end{gather*}
the antiderivative and $\tilde\phi(p)$ the Fourier transform of $\phi(x)$.
In~\cite{TanVit} the model has been studied with a~generic translation invariant product showing that the universal
properties do not depend on the particular product of the family.
In the same paper the model is formulated in the Wick--Voros matrix basis of Section~\ref{Section4.2}.
The kinetic term was computed but it was not recognized that it is of the same kind as the Grosse--Wulkenaar one, that
is, an operator of Jacobi type, while the interaction term is the same as in~\eqref{int}.
Therefore, the propagator can be found with the same techniques as in the Grosse--Wulkenhaar model.
This point deserves further investigation.
We shall come back to this issue elsewhere.

\subsection{Gauge model on the Moyal plane}\label{Section7.3}

The UV/IR mixing also occurs in gauge models on 4-dimensional Moyal space~\cite{Hayakawa, Matusis:2000jf}.
For early studies, see e.g.~\cite{gauge-var113bis, gauge-var113} and references therein.
The mixing appears in the naive noncommutative version of the Yang--Mills action given by $S_0=\frac 14\int d^4x
(F_{\mu\nu}\star F_{\mu\nu})(x)$, showing up at one-loop order as a~hard IR transverse singularity in the vacuum
polarization tensor.
Attempts to extend the Grosse--Wulkenhaar harmonic solution to a~gauge theoretic framework have singled out a~gauge
invariant action expressed as~\cite{GWW}
\begin{gather}
S_{\Omega}=\int {\mathrm d}^d x \left(\frac 14F_{\mu\nu}\star F_{\mu\nu} +\frac{\Omega^2}{4}\{\mathcal A_\mu,\mathcal
A_\nu\}^2_\star +\kappa\mathcal A_\mu\star\mathcal A_\mu\right),
\label{inducedgauge}
\end{gather}
where~$\Omega$ and~$\kappa$ are real parameters, while $\mathcal A_\mu=A_\mu-A_\mu^{\rm inv}$ is a~gauge covariant one-form
given by the dif\/ference of the gauge connection and the natural gauge invariant connection
\begin{gather*}
A_\mu^{\rm inv}= -\theta^{-1}_{\mu\nu}x^\nu.
\end{gather*}
Unfortunately, the action~\eqref{inducedgauge} is hard to deal with when it is viewed as a~functional of the gauge
potential $A_\mu$.
This is mainly due to its complicated vacuum structure explored in~\cite{GWW2}.

When expressed as a~functional of the covariant one-form ${\mathcal{A}}_\mu$, the action~\eqref{inducedgauge} bears some
similarity with a~matrix model, where the f\/ield ${\mathcal{A}}_\mu$ can be represented as an inf\/inite matrix in the Moyal
matrix base.

We will review here the two-dimensional case~\cite{MVW} and we shall consider f\/luctuations around a~particular vacuum
solution, which shall make the kinetic term of the action into a~Jacobi type operator, as in the model considered in
Section~\ref{Section7.1}.
This choice makes the model tractable and permits to invert for the propagator.

We set
\begin{gather*}
\mathcal{A}={{{\mathcal{A}}_1+i{\mathcal{A}}_2}\over{\sqrt{2}}},
\qquad
\mathcal{A}^\dag={{{\mathcal{A}}_1-i{\mathcal{A}}_2}\over{\sqrt{2}}}.
\end{gather*}
Then, one obtains
\begin{gather*}
S_\Omega[\mathcal{A}]=\int
d^2x\big(
\big(1+\Omega^2\big){\mathcal{A}}\star{\mathcal{A}^\dag}\star{\mathcal{A}}\star{\mathcal{A}}^\dag+
\big(3\Omega^2-1\big){\mathcal{A}}\star{\mathcal{A}}\star{\mathcal{A}}^\dag\star{\mathcal{A}}^\dag+
2\kappa{\mathcal{A}}\star {\mathcal{A}}^\dag
\big).
\end{gather*}
The star product used here is the Moyal star product, although any star product of the equivalence class (translation
invariant ones) would give the same results.
This action shares some similarities with the 6-vertex model although the entire analysis relies on the choice of
a~vacuum around which we shall perform f\/luctuations.

The strategy used is standard: one chooses a~particular vacuum (the background), expand the action around it, f\/ix the
background symmetry of the expanded action.

From the perspective of the present review an interesting feature of this model is the fact that, when a~particular
non-trivial vacuum is chosen, among those classif\/ied in~\cite{GWW2}, the kinetic term of the action becomes a~Jacobi
type operator, therefore invertible for the propagator in terms of orthogonal polynomials.
We therefore refer for details to~\cite{MVW} and we concentrate here on the form of the kinetic operator, when the
special vacuum is chosen.
In the Moyal basis the vacuum is expressed as
\begin{gather*}
Z(x)=\sum\limits_{m,n\in\mathbb{N}}Z_{mn}f_{mn}(x)
\end{gather*}
with
\begin{gather*}
Z_{mn}=-{{{\mathrm i}}\over{2}}{\sqrt{-3\kappa}}\delta_{m+1,n},
\qquad
\kappa<0,
\qquad
\forall\, m,n\in\mathbb{N}.
\end{gather*}
This latter expression is a~solution of the classical equation of motion for $\Omega^2={{1}\over{3}}$.
When expanded around this vacuum the kinetic part of the action becomes
\begin{gather*}
S_{\rm kin}[\phi]=\sum\limits_{m,n,k,l\in\mathbb{N}} \phi_{mn}\phi_{kl}\Delta_{mn;kl},
\end{gather*}
where $\phi= \sum\limits_{mn}\phi_{mn} f_{mn}$ are the gauge f\/ield f\/luctuations expanded in the Moyal matrix basis.
The kinetic operator reads
\begin{gather*}
\Delta^{(1/3)}_{mn;kl}=(-\kappa)(2\delta_{ml}\delta_{nk} - \delta_{k,n+1}\delta_{m,l+1}-\delta_{n,k+1}\delta_{l,m+1}),
\end{gather*}
and satisf\/ies $\Delta^{(1/3)}_{mn;kl}\ne0\iff m+n=k+l$.

The propagator, $P_{mn;kl}$, is def\/ined as in~\eqref{definvers}.
Proceeding as in the Grosse--Wulkenhaar case, we pose $\alpha=m+n=k+l$ so that
\begin{gather}
\Delta^{(1/3)}_{m,\alpha-m;\alpha-l,l}:=\Delta^{\alpha}_{m,l}=\mu^2(2\delta_{ml}- \delta_{m,l+1}-\delta_{l,m+1}),
\qquad
\forall\, m,l\in\mathbb{N},
\label{tridiagonn}
\end{gather}
where $\mu^2=-\kappa$.
Notice that in this case it does not depend on~$\alpha$.
Therefore, we set $\Delta^{\alpha}_{m,l}=\Delta_{ml}$ to simplify the notations.

One observes that $\Delta_{ml}$ is an inf\/inite real symmetric tridiagonal matrix which can be related to a~Jacobi operator.
Therefore, the diagonalization of~\eqref{tridiagonn} can be achieved by using a~suitable family of Jacobi orthogonal polynomials.

We thus arrive at the following recurrence equation
\begin{gather*}
{\mathcal{R}}_{m+1}(x)+{{\mathcal{R}}}_{m-1}(x)=(2+x){{\mathcal{R}}}_{m}(x),
\qquad
\forall\, m\in\mathbb{N},
\end{gather*}
where we have posed $x=-\lambda_q$.
On restricting to $N\times N$ submatrices, it is possible to show that the recurrence equation above is satisf\/ied~by
Chebyschev polynomials of second kind~\cite{kks}:
\begin{gather*}
U_m(t):=(m+1)\,{}_2F_1\left(-m,m+2;{{3}\over{2}};
{{1-t}\over{2}}\right),
\qquad
\forall\, m\in\mathbb{N},
\end{gather*}
where ${}_2F_1$ denotes the hypergeometric function.
Moreover, the eigenvalues are exactly given by the roots of ${{\mathcal{R}}}_N(x)$.
So we have
\begin{gather*}
{\mathcal{R}}_m(x)=f(x)U_m\left(\frac{2+x}{2}\right),
\qquad
\forall\, m\in\mathbb{N},
\end{gather*}
where $f(x)$ is a~normalization function to be determined by the orthonormality condition~\eqref{orthogpoly}.
The eigenvalues of $\Delta^N_{ml}$ are now entirely determined by the roots of $U_N(t)$.
These are given by $t_k^N=\cos({{(k+1)\pi}\over{N+1}})$, $k=0,2,\dots,N-1$.
Then, the eigenvalues for the kinetic operator $\Delta^N_{ml}$ are
\begin{gather*}
\mu^2\lambda_k^N=2\mu^2\left(1-\cos\left({{(k+1)\pi}\over{N+1}}\right)\right),
\qquad
k\in\{0,2,\dots,N-1\},
\end{gather*}
and satisfy for f\/inite~$N$
\begin{gather*}
0<\mu^2\lambda_N^k <4\mu^2.
\end{gather*}
Thus, we have obtained:
\begin{gather}
{\mathcal{R}}^N_{mq}=f(N,q)U_m\big(t^N_q\big)=f(N,q){{\sin\big[{{\pi(m+1)(q+1)}\over{N+1}}\big]}\over{\sin\big[{{\pi(q+1)}\over{N+1}}\big]}},
\qquad
0\le m,
\quad
q\le N-1,
\label{finalrml}
\end{gather}
where we used $U_m(\cos\theta)={{\sin((m+1)\theta)}\over{\sin\theta}}$.

The normalization function is found to be
\begin{gather*}
f(N,m)=\left((-1)^m(N+1){{\sin\big[{{N(m+1)\pi}\over{N+1}}\big]}\over {\sin^3\big[{{(m+1)\pi}\over{N+1}}\big]}}\right)^{-{{1}\over{2}}},
\qquad
0\le p,
\quad
m\le N-1.
\end{gather*}
Once we have the polynomials which diagonalize the kinetic term we can invert for the propagator.
Keeping in mind equations~\eqref{definvers} and~\eqref{tridiagonn}, we set $P_{mn}:=P_{m,\alpha-n;\alpha-l,l}$ where
$\alpha=m+n=k+l$.
It follows from the above that for f\/ixed~$N$ the inverse of $\Delta^N_{mn}$ denoted by $P^N_{mn}$ can be written as
\begin{gather}
P^N_{mn}={{1}\over{2\mu^2}}\sum\limits_{p=0}^{N-1}f^2(N,p)U_m\big(t^N_p\big){{1}\over{1-t^N_p}}U_n\big(t^N_p\big).
\label{propagfinite2}
\end{gather}
Taking the limit $N\to\infty$, the comparison of the relation
$\delta_{ml}=\sum\limits_p{{\mathcal{R}}}^N_{mp}{{\mathcal{R}}}^N_{lp}$ where the ${{\mathcal{R}}}^N_{mn}$'s are given~by
equation~\eqref{finalrml} to the orthogonality relation among the Chebyshev polynomials $U_n$
\begin{gather*}
\int_{-1}^1{\rm d}\mu(x)U_m(x)U_n(x)={{\pi}\over{2}}\delta_{mn},
\qquad
{\rm d}\mu(x)={\rm d}x{\sqrt{1-x^2}},
\end{gather*}
permits one to trade the factor $f^2(N,p)$ in $P^N_{mn}$~\eqref{propagfinite2} for the compactly supported integration
measure $d\mu(x)$.

We f\/inally obtain the following rather simple expression for the inverse of the kinetic operator~\eqref{tridiagonn}
\begin{gather}
P_{mn;kl} = \delta_{m+n,k+l}P_{ml},
\qquad
P_{ml} = {{1}\over{\pi\mu^2}}\int_{-1}^1{\rm d}x
\sqrt{{{1+x}\over{1-x}}}U_m(x)U_l(x).
\label{propag-fin1}
\end{gather}
Notice that the integral in~\eqref{propag-fin1} is well-def\/ined leading to f\/inite $P_{ml}$ when~$m$ and~$l$ are f\/inite.

\subsection[The scalar model on $\mathbb R^3_\lambda$]{The scalar model on $\boldsymbol{\mathbb R^3_\lambda}$} \label{Section7.4}

In this section we review a~family of scalar f\/ield theories on ${\mathbb R}^3_\lambda$, the noncommutative algebra
introduced in Section~\ref{starprodred}.
The contents and presentation are based on~\cite{VW}, where one loop calculations were performed.
Here the stress will be, as for the models described in the previous sections, on the use of a~matrix basis (in this
case the one of Section~\ref{R3lambda}) to obtain a~non-local matrix model and show that the kinetic term of the theory
is of Jacobi type, so that it can be inverted for the propagator using standard techniques of orthogonal polynomials.

Let us recall that the algebra ${\mathbb R}^3_\lambda$ is generated by the coordinate functions $x_\mu$, $\mu=0,\dots,3$.
The coordinate $x_0$ is in the center of the algebra and plays the role of the radius of fuzzy two-spheres which foliate
the whole algebra.

Let
\begin{gather}
S[\phi]=\int \phi\star\big(\Delta+\mu^2\big) \phi + \frac{g}{4!}\phi\star\phi\star\phi\star\phi,
\label{action}
\end{gather}
where~$\Delta$ is the Laplacian def\/ined as
\begin{gather}
\Delta \phi = \alpha \sum\limits_i D_i^2\phi + \frac{\beta}{\kappa^4}x_0\star x_0\star\phi
\label{lapl}
\end{gather}
and
\begin{gather}
D_i=\kappa^{-2}[x_i,
\,
\cdot
\,
]_\star,
\qquad
i=1,\dots,3
\label{innerd}
\end{gather}
are inner derivations of ${\mathbb R}^3_\lambda$.
The mass dimensions are $[\phi]=\frac{1}{2}$, $[g]=1$, $[D_i]=1$.
$\alpha$ and~$\beta$ are dimensionless parameters.

The second term in the Laplacian has been added in order to introduce radial dynamics.
From~\eqref{starsu2} we have indeed
\begin{gather*}
[x_i,\phi]_\star= - i\lambda \epsilon_{ijk}x_j \partial_k \phi
\end{gather*}
so that the f\/irst term, that is $[x_i,[x_i,\phi]_\star]\star$ can only reproduce tangent dynamics on fuzzy spheres; this
is indeed the Laplacian usually introduced for quantum f\/ield theories on the fuzzy sphere (cf.\
Section~\ref{Section8.2}).
Whereas{\samepage
\begin{gather*}
x_0\star \phi = x_0 \phi + \frac{\lambda}{2} x_i \partial_i \phi
\end{gather*}
contains the dilation operator in the radial direction.}

Therefore, the highest derivative term of the Laplacian def\/ined in~\eqref{lapl} can be made into the ordinary Laplacian
on ${\mathbb R}^3$ multiplied by $x_0^2$, for the parameters~$\alpha$ and~$\beta$ appropriately chosen.

For simplicity, we restrict the analysis to $\alpha$, $\beta$ positive, which is a~suf\/f\/icient condition for the spectrum to be positive.

It is not dif\/f\/icult to verify that the following relations old
\begin{gather*}
[x_+,[x_-, v^j_{m\tilde m}]_\star]_\star = \lambda^2 \big\{\big((j+m)(j-m+1)+ (j+\tilde m+1) (j-\tilde m) \big)
v^j_{m\tilde m}
\\
\hphantom{[x_+,[x_-, v^j_{m\tilde m}]_\star]_\star=}{}
 -\sqrt{(j+m)(j-m+1)(j+\tilde m) (j-\tilde m +1)} v^j_{m-1 \tilde m-1}
\\
\hphantom{[x_+,[x_-, v^j_{m\tilde m}]_\star]_\star=}{}
 - \sqrt{(j+m+1)(j-m)(j+\tilde m+1) (j-\tilde m)} v^j_{m+1 \tilde m+1} \big\},
\\
[x_-,[x_+, v^j_{m\tilde m}]_\star]_\star = \lambda^2 \big\{\big((j+m+1)(j-m)+ (j+\tilde m) (j-\tilde m+1) \big)
v^j_{m\tilde m}
\\
\hphantom{[x_-,[x_+, v^j_{m\tilde m}]_\star]_\star =}{}
 -\sqrt{(j+m)(j-m+1)(j+\tilde m) (j-\tilde m+1)} v^j_{m-1 \tilde m-1}
\\
\hphantom{[x_-,[x_+, v^j_{m\tilde m}]_\star]_\star =}{}
 - \sqrt{(j+m+1)(j-m)(j+\tilde m+1) (j-\tilde m)} v^j_{m+1 \tilde m+1} \big\},
\\
\big[x_3,\big[x_3, v^j_{m\tilde m}\big]_\star\big]_\star = \lambda^2 (m-\tilde m)^2 v^j_{m\tilde m},
\\
x_0\star x_0 \star v^j_{m\tilde m} = \lambda^2 j^2 v^j_{m\tilde m}.
\end{gather*}
On expanding the f\/ields in the matrix basis$ \phi=\sum\limits_{j,m\tilde m} \phi^j_{m\tilde m} v^j_{m \tilde m}$ we
rewrite the action in~\eqref{action} as a~matrix model action
\begin{gather}
S[\phi]= \kappa^3\big\{\tr(\Phi(\Delta(\alpha,\beta))\Phi)+{{g}\over{4!}}\tr(\Phi\Phi\Phi\Phi)\big\},
\label{action1}
\end{gather}
where sums are understood over all the indices and $\tr:=\sum\limits_j \tr_j$.
The kinetic operator may be computed to be
\begin{gather*}
(\Delta(\alpha,\beta))^{j_1 j_2}_{m_1\tilde m_1;m_2\tilde m_2} = \frac{1}{\pi^2\theta^2} \int v^{j_1}_{m_1 \tilde
m_1}\star(\Delta (\alpha,\beta)) v^{j_2}_{m_2\tilde m_2}
\\
\hphantom{(\Delta(\alpha,\beta))^{j_1 j_2}_{m_1\tilde m_1;m_2\tilde m_2}}{}
 = \frac{\lambda^2}{\kappa^4}\delta^{j_1 j_2}\big\{\delta_{\tilde m_1 m_2}\delta_{m_1 \tilde m_2} D^{j_2}_{m_2\tilde
m_2}-\delta_{\tilde m_1, m_2+1}\delta_{m_1, \tilde m_2+1}B^{j_2}_{m_2,\tilde m_2}
\\
\hphantom{(\Delta(\alpha,\beta))^{j_1 j_2}_{m_1\tilde m_1;m_2\tilde m_2}=}{}
 -\delta_{\tilde m_1, m_2-1}\delta_{m_1, \tilde m_2-1} H^{j_2}_{m_2,\tilde m_2}\big\}
\end{gather*}
with
\begin{gather*}
D^j_{m_2\tilde m_2} = \big[({2\alpha}+{\beta})j^2 + {2\alpha}(j_2- m_2 \tilde m_2)\big] +\lambda^2{\mu^2},
\\
B^j_{m_2\tilde m_2} = \alpha \sqrt{(j+m_2+1)(j-m_2)(j+\tilde m_2+1)(j-\tilde m_2)},
\\
H^j_{m_2\tilde m_2} = \alpha\sqrt{(j+m_2)(j-m_2+1)(j+\tilde m_2)(j-\tilde m_2+1)}.
\end{gather*}
Let us notice that the use of the matrix basis $v^j_{mn}$ yields an interaction term which is diagonal whereas the
kinetic term is not diagonal.
Had we used the expansion of~$\phi$ in the fuzzy harmonics base $(Y^j_{lk})$, $j\in{{\mathbb{N}}\over{2}}$,
$l\in\mathbb{N}$, $0\le l\le 2j$, $-l\le k\le l$ (see Section~\ref{Section8.2}), we would have obtained a~diagonal
kinetic term with a~non-diagonal interaction term.
The latter will be the choice in Section~\ref{Section8.2} where we follow the traditional approach to the study of the
fuzzy sphere Laplacian.

Moreover, we observe that the action~\eqref{action1} is expressed as an inf\/inite sum of contributions, namely
$S[\Phi]=\sum\limits_{j\in{{\mathbb{N}}\over{2}}}S^{(j)}[\Phi]$, where the expression for $S^{(j)}$ can be read of\/f
from~\eqref{action1} and describes a~scalar action on the fuzzy sphere $\mathbb{S}^j$.

We now pass to the calculation of the propagator, through the inversion of the kinetic term in the action.
Because of the remark above, this is expressible into a~block diagonal form.
Explicitly
\begin{gather*}
S_{Kin}[\Phi]=\kappa^3\sum\limits_{j} \sum\limits_{m,\tilde m} \phi^{j_1}_{m_1 \tilde m_1}(\Delta)^{j_1 j_2}_{m_1\tilde
m_1;m_2\tilde m_2}\phi^{j_2}_{m_2\tilde m_2}.
\end{gather*}
Since the mass term is diagonal, let us put it to zero for the moment.
We shall restore it at the end.
One has the following law of indices conservation
\begin{gather*}
\Delta^{j_1 j_2}_{mn;kl}\ne 0 \implies j_1=j_2,
\quad
m+k= n+l.
\end{gather*}
The inverse of $\Delta^{j_1 j_2}_{mn;kl}(\alpha,\beta)$ is thus def\/ined~by
\begin{gather*}
\sum\limits_{k,l=-j_2}^{j_2}\Delta^{j_1 j_2}_{mn;lk}P^{j_2 j_3}_{lk;rs}=\delta^{j_1 j_3}\delta_{ms}\delta_{nr},
\qquad
\sum\limits_{m,n=-j_2}^{j_2}P^{j_1 j_2}_{rs;mn}\Delta^{j_2 j_3}_{mn;kl}=\delta^{j_1 j_3}\delta_{rl}\delta_{sk},
\end{gather*}
for which the law of indices conservation still holds true as
\begin{gather*}
P^{j_1 j_2}_{mn;kl}\ne0\implies j_1=j_2,
\ \
m+k= n+l.
\end{gather*}
To determine $P^{j_1 j_2}_{mn;kl}$ one has to diagonalize $\Delta^{j_1 j_2}_{mn;kl}$ along the same lines as in previous
sections, by means of orthogonal polynomials.
This is done in detail in~\cite{VW} where the orthogonal polynomials are found to be the dual Hahn polynomials.
Here however we take a~shortcut, because we already know an alternative orthogonal basis for ${\mathbb R}^3_\lambda$
where the kinetic part of the action is diagonal, that is the fuzzy spherical harmonics.
It can be shown that dual Hahn polynomials and the fuzzy spherical harmonics are indeed the same object, up to
a~proportionality factor.

\subsubsection{The kinetic action in the fuzzy spherical harmonics base} \label{Section7.4.1}

Fuzzy Spherical Harmonics Operators, are, up to normalization factors, irreducible tensor ope\-ra\-tors
\begin{gather*}
\hat Y^j_{lk}\in \operatorname{End}({\mathcal{V}}^j),
\qquad
l\in\mathbb{N},
\qquad
0\le l\le 2j,
\qquad
-l\le k\le l,
\end{gather*}
whereas the unhatted objects $Y^j_{lk}$ are their symbols and are sometimes referred to as fuzzy spherical harmonics
with no other specif\/ication (notice however that the functional form of the symbols does depend on the dequantization
map that has been chosen).
Concerning the def\/inition and normalization of the fuzzy spherical harmonics operators, we use the following
conventions.
We set
\begin{gather*}
J_{\pm}=\frac{{\hat x}_{\pm}}{\lambda}.
\end{gather*}
We have, for $l=m$,
\begin{gather*}
\hat Y^j_{ll}:=(-1)^l \frac{\sqrt{2j+1}}{l !} \frac {\sqrt{(2l+1)! (2j-l)!}} {(2j+l+1)!} (J_+)^l
\end{gather*}
while the others are def\/ined recursively through the action of $J_-$
\begin{gather*}
\hat Y^j_{lk}:= [(l+k+1) (l-k)]^{-\frac{1}{2}} [J_-,\hat Y^j_{l,k+1}],
\end{gather*}
and satisfy
\begin{gather*}
\big(\hat Y^j_{lk}\big)^\dag=(-1)^{k-2j}\hat Y^j_{l,-k},
\qquad
\langle\hat Y^j_{l_1 k_1},
\hat Y^j_{l_2 k_2}\rangle=\tr\big(\big(\hat Y^j_{l_1k_1}\big)^\dag{\hat Y}^j_{l_2k_2}\big)=(2j+1)\delta_{l_1l_2}\delta_{k_1k_2}.
\end{gather*}
The symbols are def\/ined through the dequantization map~\eqref{wigner}
\begin{gather}
Y^j_{lk}:=\langle z|\,\hat Y^j_{lk}\,|z\rangle.
\label{fuzzyha}
\end{gather}
We have then
\begin{gather*}
[x_i,[x_i, Y^j_{lk}]_\star ]_\star=\lambda^2 \langle z|[ J_{i},[ J_{i},{\hat Y}^j_{lk} ] ]
|z\rangle=\lambda^2
l(l+1)
Y^{j}_{l k}.
\end{gather*}
In order to evaluate the action of the full Laplacian~\eqref{lapl} on the fuzzy spherical harmonics we need to compute
$x_0\star Y^j_{lk}$.
To this we express the fuzzy spherical harmonics in the canonical base~$v^j_{m\tilde m}$
\begin{gather*}
Y^j_{lk}=\sum\limits_{-j\le m,\tilde m\le j} \big(Y^j_{lk}\big)_{m \tilde m} v^j_{m \tilde m},
\end{gather*}
where the coef\/f\/icients are given in terms of Clebsch--Gordan coef\/f\/icients~by
\begin{gather*}
(Y^j_{lk})_{m \tilde m}=\langle \hat v^j_{m \tilde m}|\hat Y^j_{lk}\rangle={\sqrt{2j+1}}(-1)^{j-\tilde m}\cg{j}{m}{j}{-\tilde m}{l}{k},
\qquad
-j\le m,
\qquad
\tilde m\le j,
\\
\big({Y^j_{lk}}^\dag\big)_{m \tilde m}=(-1)^{-2j} \big(Y^j_{lk}\big)_{\tilde m m}.
\end{gather*}
We have then
\begin{gather*}
x_0\star Y^j_{lk}= \sum\limits_{-j\le m,\tilde m\le j} \big(Y^j_{lk}\big)_{m \tilde m}
x_0\star v^j_{m \tilde m}= \lambda j Y^j_{lk}.
\end{gather*}
Thus we verify that in the fuzzy spherical harmonics base the whole kinetic term is diagonal,
\begin{gather*}
\Delta (\alpha,\beta) Y^j_{lk}= \frac{\lambda^2}{\kappa^4}\big(\alpha l(l+1)+\beta j^2 \big)Y^j_{lk},
\qquad
j\in{{\mathbb{N}}\over{2}},
\quad
0\le l\le 2j,
\quad
l\in\mathbb{N},
\quad
-l\le k\le l,
\end{gather*}
with eigenvalues
\begin{gather*}
\frac{\lambda^2}{\kappa^4}\gamma(j,l;\alpha,\beta):= \frac{\lambda^2}{\kappa^4}\big(\alpha l (l+1)+\beta j^2 \big).
\end{gather*}
We can expand the f\/ields $\phi \in {\mathbb R}^3_\lambda$ in the fuzzy harmonics base $\phi=\sum\limits_{j\in
{{{{\mathbb{N}}}\over{2}}}}\sum\limits_{l=0}^{2j}\sum\limits_{k=-l}^l\varphi^j_{lk} Y^j_{lk}$, with the coef\/f\/icients
$\varphi^j_{lk}$ related to those in the canonical base $\phi^j_{m\tilde m}$ by a~change of basis.

Therefore, we can compute the kinetic action in the fuzzy harmonics base to be
\begin{gather*}
\int \phi\star (\Delta+\mu^2) \phi= \frac{\lambda^2}{\kappa} \sum | \varphi^{j}_{l k}|^2 (2j+1)
\left(\gamma(j,l;\alpha,\beta)+\frac{\kappa^4}{\lambda^2}\mu^2\right),
\end{gather*}
which is positive for $\alpha, \beta \ge0$.
We def\/ine for further convenience
\begin{gather*}
(\Delta_{\rm diag})^{j_1 j_2}_{l_1 k_1 l_2 k_2}= \frac{1}{\lambda^3}\int Y^{j_1}_{l_1 k_1} \star \Delta(\alpha,\beta)Y^{j_2}_{l_2 k_2}
\\
\phantom{(\Delta_{\rm diag})^{j_1 j_2}_{l_1 k_1 l_2 k_2}}
= \frac{1}{\lambda^2} (-1)^{k_1+2j_1} (2j_1+1) \gamma(j_1,l_1;\alpha,\beta) \delta^{j_1 j_2}\delta_{l_1 l_2} \delta_{-k_1 k_2}.
\end{gather*}
Then, the kinetic term in the canonical basis may be expressed in terms of the diagonal one
\begin{gather*}
\Delta^{j_1 j_2}_{m_1\tilde m_1 m_2\tilde m_2}
=\frac{1}{(2 j_1+1)^2}\big(Y^{j_1}_{l_1 k_1}\big)_{m_1\tilde m_1} \big(\Delta^{j_1j_2}_{\rm diag}\big)_{l_1 k_1 l_2 k_2}
\big(Y^{j_2}_{l_2 k_2}\big)_{m_2\tilde m_2}.
\end{gather*}
The propagator is then
\begin{gather*}
[P^{j_1 j_2}
]_{m_1\tilde m_1 m_2\tilde m_2}
=\big(Y^{j_1}_{l_1 k_1}\big)_{m_1\tilde m_1}\big[\big(\Delta_{\rm diag}^{j_1 j_2}\big)^{-1}\big]_{l_1k_1 l_2 k_2}
\big(Y^{j_2}_{l_2 k_2}\big)_{m_2\tilde m_2}.
\end{gather*}
On replacing the expression for the diagonal inverse we f\/inally obtain
\begin{gather}
[P^{j_1 j_2}]_{m_1\tilde m_1 m_2\tilde m_2} = (-1)^{-k+2j_1} \delta^{j_1 j_2}
\nonumber
\\
\hphantom{[P^{j_1 j_2}]_{m_1\tilde m_1 m_2\tilde m_2} =}{}
\times
\sum\limits_{l=0}^{2j_1}
\sum\limits_{k=-l}^l \frac{\kappa^4}{\lambda^2}\frac{1}{(2j_1\!+1) \big(\gamma(j_1,l; \alpha,\beta)+\mu^2\big)}
\big(Y^{j_1\dag}_{l k}\big)_{m_1\tilde m_1} \big(Y^{j_2}_{l k}\big)_{m_2\tilde m_2}.\!\!\!
\label{propR3}
\end{gather}
In~\cite{VW} one loop calculations have been performed showing the absence of divergences.

We f\/inally that these results may be generalized to gauge theories on ${\mathbb R}^3_\lambda$.
In~\cite{GVW} the following gauge model has been considered
\begin{gather}
S_{\rm cl}(A_i)=\tr\big(\alpha{\mathcal{A}}_i{\mathcal{A}}_j {\mathcal{A}}_j {\mathcal{A}}_i+\beta{\mathcal{A}}_i {\mathcal{A}}_j {\mathcal{A}}_i
{\mathcal{A}}_j+\Theta\varepsilon_{ijk}{\mathcal{A}}_i {\mathcal{A}}_j {\mathcal{A}}_k+m{\mathcal{A}}_i {\mathcal{A}}_i \big),
\label{spectral-action}
\end{gather}
where ${\mathcal{A}}_i=-iA_i+\eta_i$ and~$\alpha$,~$\beta$,~$\Theta$,~$m$ are real parameters.
$A_i$ is the gauge potential and $\eta_i$ is the invariant connection associated to the dif\/ferential calculus on
${\mathbb R}^3_\lambda$
\begin{gather*}
\eta(D_i):=\eta_i=\frac{i}{\kappa^2}x_i,
\end{gather*}
where $D_i$ are the inner derivations of the algebra introduced in equation~\eqref{innerd}.
By requiring that no linear terms in $A_i$ be involved, the action~\eqref{spectral-action} may be rewritten as
\begin{gather*}
S_{\rm cl}(A_i)=\tr\left(F^\dag_{ij}F_{ij}+\gamma(\epsilon_{ijk}
{\mathcal{A}}_i{\mathcal{A}}_j{\mathcal{A}}_k+\frac{3}{2}\frac{\lambda}{\kappa^2}{\mathcal{A}}_i{\mathcal{A}}_i)\right).
\end{gather*}
with appropriately def\/ined parameters.
The total action is thus rewritten as the sum of a~Yang--Mills and a~Chern--Simons term,
\begin{gather*}
S_{\rm cl}(A_i)=\tr F^\dag_{ij}F_{ij}+S^{\rm CS}_{\rm cl}(A_i),
\end{gather*}
which is of the same form as the Alekseev--Recknagel--Schomerus gauge action on the fuzzy sphere~\cite{ARS00}, although
here we have a~sum over all fuzzy spheres of the foliation of~${\mathbb R}^3_\lambda$.

The model has been studied in the matrix basis of ${\mathbb R}^3_\lambda$ showing that, when the action is formally
massless, the gauge and ghost propagators are of the same form as the scalar pro\-pa\-ga\-tor~\eqref{propR3} found above.
This result has been used to perform one loop calculations.
It is found that the infrared singularity of the propagator stemming from masslessness disappears from the computation
of the correlation functions.
Moreover it is shown that this massless gauge invariant model on $\mathbb{R}^3_\lambda$ has quantum instabilities of the
vacuum, signaled by the occurrence of non vanishing tadpole (1-point) functions for some but not all of the components
of the gauge potential.

We close this section observing that all the models considered are connected to Jacobi type kinetic operators, which
give rise to three term recurrence equations.
These are solved for specif\/ic families of polynomials which allow in turn to determine the propagator.
A~systematic analysis is performed in~\cite{GW14}.
However, it is interesting to notice that f\/ive terms recurrence relations emerge, for example in the two-dimensional
gauge model, with a~dif\/ferent choice for the vacuum, but also in the three-dimensional gauge model on ${\mathbb
R}^3_\lambda$, in the massive case, which might be worth to investigate.

\section{Fuzzy spaces} \label{Section8}

Fuzzy spaces are matrix approximations of ordinary spaces.
Their importance lies in the fact that, although the algebra which approximates the original functions on the space is
f\/inite-dimensional, the original group of isometries is preserved.
The literature on fuzzy spaces is vast (see~\cite{BalSachinSeczkin, Madorebook}
and references therein),
in this section we will limit ourselves to a~presentation of these fuzzy space which shows how they can be interpreted
in a~way similar to the matrix basis for the $\star$ products in the previous section.

\subsection{The fuzzy torus} \label{Section8.1}

The fuzzy torus is a~f\/inite-dimensional of the noncommutative torus~\cite{Rieffeltorus}, which is probably the most
studied noncommutative space.
It is in some sense a~compact version of the Moyal plane introduced in the previous section.
Consider the algebra of functions on a~two-dimensional torus\footnote{Higher-dimensional cases can be studied, but in
this review we will conf\/ine ourselves to two dimensions.}.
In Fourier transform they can be represented as
\begin{gather}
f=\sum\limits_{n_1,n_2=-\infty}^\infty f_{n_1n_2} {\mathrm e}^{2\pi{\mathrm i} n_1 x_1}{\mathrm e}^{2\pi{\mathrm i} n_2x_2},
\label{torusexpansion}
\end{gather}
where we impose that the $f_{n_1n_2}$ decrease exponentially as $n_i\to\pm\infty$.
The noncommutative torus is obtained with the substitution ${\mathrm e}^{2\pi{\mathrm i} x_i}\to U_i$ with the
condition
\begin{gather}
U_1U_2={\mathrm e}^{2\pi{\mathrm i}\theta}U_2U_1.
\label{nctorus}
\end{gather}
Loosely speaking this is what would be obtained imposing the commutation relation~\eqref{xcomm}, except that of course
the~$x$'s are not well def\/ined quantities on a~torus.
One can represent the $U_i$'s as operators on the Hilbert space of $L_2(S^1)$ functions a~circle as follows:
\begin{gather*}
U_1\psi(\alpha)=\psi(\alpha+2\pi\theta),
\qquad
U_2\psi(\alpha)={\mathrm e}^{2\pi\alpha}\psi(\alpha).
\end{gather*}
It is immediate to verify relation~\eqref{nctorus}.
It follows that the noncommutative torus is given by a~Weyl map
\begin{gather*}
\hat\Omega(f)=\sum\limits_{n_1,n_2=-\infty}^\infty f_{n_1n_2} U_1^{n_1}U_2^{n_2}
\end{gather*}
giving rise to the noncommutative $\star$ product def\/ined by the twisted convolution of Fourier coef\/f\/icients:
\begin{gather*}
(f*g)_{n_1 n_2}=\sum\limits_{m_1,m_2=-\infty}^\infty f_{m_1m_2} g_{n_1-m_1, n_2-m_2} {\mathrm e}^{2\pi{\mathrm
i}(n_1m_2-n_2m_1)}.
\end{gather*}
The representation of the operators $U_1$ and $U_2$ in the discrete basis of $L_2(S^1)$ given by $\varphi_p={\mathrm
e}^{2\pi{\mathrm i} p\alpha}$ is given~by
\begin{gather*}
{U_1}_{pq}=\delta_{p,q-1},
\qquad
{U_2}_{pq}={\mathrm e}^{2\pi p}\delta_{p,q}
\qquad
(\text{no sum over $p$}).
\end{gather*}

In the rational case of $\theta=\frac MN$ it is possible to f\/ind a~f\/inite $(N\times N)$-dimensional representation of the
$U_i$'s as:
\begin{gather*}
U_1^{(N)} =  \left({
\begin{matrix}
1& & & &
\\
&{\mathrm e}^{2\pi{\mathrm i}\theta}& & &
\\
& &{\mathrm e}^{2\pi{\mathrm i}\theta}& &
\\
& & &\ddots&
\\
& & & & {\mathrm e}^{2\pi{\mathrm i}(N-1)\theta}
\end{matrix}
}\right),
\qquad
U_2^{(N)} = \left(
\begin{matrix}
0&1& & &0
\\
&0&1& &
\\
& &\ddots&\ddots&
\\
& & &\ddots&1
\\
1& & & &0
\end{matrix}
\right).
\end{gather*}
They are unitary and traceless (since
$\sum\limits_{k=0}^{N-1} {{\mathrm e}^{2\pi{\mathrm i}\theta}}^k = 0$), satisfy
\begin{gather*}
U_i^M = \mathbb I_N
\end{gather*}
and obey the commutation relation~\eqref{nctorus}.
The algebra generated by the $U_i^{(N)}$ with the expansion~\eqref{torusexpansion} is called a~\emph{fuzzy torus}.
Note that a~noncommutative torus with rational $\theta=\frac MN$ is not the same algebra of a~fuzzy torus with the
same~$\theta$.
The former is inf\/inite-dimensional, while the latter if f\/inite-dimensional.
The fuzzy torus can however be seen an an approximation of a~noncommutative torus, by taking the value of~$N$, and hence
the size of the matrices larger and larger.
Since any irrational number can be approximated arbitrarily by a~sequence of rationals, for example using continuous
fractions, there is a~sequence of f\/inite-dimensional algebras which approximate the algebra of the noncommutative torus.
The appropriate tool for this approximation is the inductive limit~\cite{bratteli}, but the inf\/inite-dimensional algebra
of the noncommutative (or commutative) torus is not approximable by a~sequence of f\/inite-dimensional algebras.
It is however possible~\cite{LandiLizziSzaboNCtorus1,PimsnerVoiculescu} to prove that the inductive limit of a~sequence
of these f\/inite-dimensional algebras converges to a~larger algebra which contains the noncommutative tours, as well as
the algebra of all the tori which are Morita equivalent to it.

\subsection{The fuzzy sphere} \label{Section8.2}

The fuzzy sphere~\cite{Hoppethesis, Madore} is the most famous example of fuzzy space, and is usually presented using
the identif\/ication $x_i\propto J_i$, where the~$x$'s are the coordinates on ${\mathbb R}^3$ and the~$J$'s the generators
of angular momentum in a~particular representation.
The sphere constraint $\sum\limits_i x_i^2=R^2$ is then equivalent to the Casimir relation $\sum\limits_i J_i^2\propto
\mathbb I_{2j+1}$.
In this section we will present the fuzzy sphere as an example of Weyl--Wigner correspondence and an instance of a~star
product.
The product is based, as in the case of the Wick--Voros plane, on the use of coherent states.
Since the sphere is a~coadjoint orbit of ${\rm SU}(2)$ the relevant coherent states are the generalization of the usual ones
pertaining to the one related this group.
Notice however, that it could be equivalently considered as a~noncommutative subalgebra of ${\mathbb R}^3_\lambda$ (cf.\
Section~\ref{R3lambda} at some f\/ixed value of~$x_0$).

Consider ${\rm SU}(2)$ in a~particular representation.
The construction can be made for every Lie algebra~\cite{Perelomov}, and related coadjoint orbits.
Consider a~representation of the group on the f\/inite-dimensional Hilbert space $\mathcal H_{2j+1}$:
\begin{gather*}
g\in\mathrm{SU}(2)\to U(g),
\end{gather*}
where $U(g)$ is a~$2j+1\times 2j +1$ matrix, $j\in\mathbb Z/2$.
Consider a~vector $\ket\psi$.
A~subgroup $\mathrm H_\psi\subset\mathrm{SU}(2)$ will leave it invariant up to a~phase.
Consider now a~f\/iducial vector $\ket{\psi_0}$ such that $\mathrm H_{\psi_0}$ is maximal.
A~natural choice for the f\/iducial state is a~highest weight vector for the representation $\ket{\psi_0}=\ket{j,j}$,
where we use the basis $\ket{j,m}$ of simultaneous eigenvectors of $J^2$ and $J_3$, with $m=-j, -j+1,\ldots, j$.
The sphere is the quotient of ${\rm SU}(2)$, which topologically is a~three sphere~$S^3$, by the subgroup~$\mathrm H_{\psi_0}$,
which in this case is ${\rm U}(1)$,
\begin{gather*}
S^2=\mathrm{SU}(2)/{\mathrm H}_{\psi_0}.
\end{gather*}

Consider the usual basis of $S^3$ given by the three Euler angles $\alpha\in[0,4\pi)$, $\beta\in[0,\pi)$,
$\gamma\in[0,2\pi)$.
the corresponding element in ${\rm SU}(2)$ is given~by
\begin{gather*}
U(\alpha,\beta,\gamma)={\mathrm e}^{-{\mathrm i}\alpha J_3}{\mathrm e}^{-{\mathrm i}\beta J_2}{\mathrm e}^{-{\mathrm
i}\gamma J_3}.
\end{gather*}
The points for which $\beta=0$ are left invariant up to a~phase.
The sphere $S^2$ can therefore be characterized by the coordinates~$\beta$ and $\alpha\; {\rm mod}\; 2\pi$,
which we may identify with the usual coordinates on the sphere~$\vartheta$ and~$\varphi$.

Choosing a~representative element~$g$ in each equivalence class of the quotient, the set of coherent states is def\/ined~by
\begin{gather*}
|\vartheta,\varphi\rangle_{2j+1}=U(g)|j,j\rangle.
\end{gather*}
They depend on the dimension of the representation.
Projecting onto the basis elements $\ket{j,m}$ one f\/inds
\begin{gather*}
|\vartheta,\varphi\rangle_{2j+1} = \sum\limits_{m=-j}^{j}
 \sqrt{\frac{(2j)!}{(j+m)!(j-m)!}}  \cos^{j+m} \frac\vartheta2
\sin^{j-m}\frac{\vartheta}{2} {\mathrm e}^{-{\mathrm i} m\varphi}\,|j,m\rangle,
\end{gather*}
As in the earlier case, coherent states are non-orthogonal and overcomplete
\begin{gather*}
{}_{2j+1}\langle\vartheta^{\prime},\varphi^{\prime}|\vartheta,\varphi\rangle_{2j+1} = {\mathrm e}^{-{\mathrm i}
j(\varphi^{\prime}-\varphi)}  \left[{\mathrm e}^{{\mathrm i}(\varphi^{\prime}-\varphi)}\cos
\frac{\vartheta}{2} \cos\frac{\vartheta^{\prime}}{2}  +\sin
\frac{\vartheta}{2}\sin\frac{\vartheta^{\prime}}{2} \right]^{2j},
\nonumber
\\
\mathbb I =\frac{2j+1}{4\pi}\int_{S^{2}} {\mathrm d}\Omega|\vartheta,
\varphi\rangle_{2j+1}\,{}_{2j+1}\langle\vartheta,\varphi|,
\end{gather*}
where ${\mathrm d}\Omega=\sin\vartheta   {\mathrm d}\vartheta  {\mathrm d}\varphi$.

As in the case discussed in Section~\eqref{Section3.3} we can use coherent states to def\/ine a~map from operators to functions.
Note that in this case, as in the fuzzy torus case, the map is not one-to-one.
\begin{gather}
\hat{F}^{(2j+1)} \in \mathbb{M}_{2j+1}(\mathbb C)
\
\longmapsto
\
f^{(2j+1)} \in C\big(S^{2}\big),
\nonumber
\\
f^{\left(2j+1\right)}(\vartheta,\varphi)=
{}_{2j+1}\langle\vartheta,\varphi|\hat{F}^{(2j+1)}|\vartheta,\varphi\rangle_{2j+1}.
\label{berezinsymsph}
\end{gather}
This is also called the Berezin symbol of the matrix~\cite{Berezin}.

Spherical harmonics operators, already introduced in Section~\ref{Section7.4.1}, form a~basis for the algebra of
${2j+1}\times {2j+1}$ matrices.
Therefore elements $\hat{F}^{({2j+1})}\in\mathbb{M}_{{2j+1}}(\mathbb C)$ can be expanded as
\begin{gather*}
\hat{F}^{(2j+1)}= \sum\limits_{l=0}^{j}\sum\limits_{m=-l}^{l} 
F^{(2j+1)}_{lm} \hat{Y}^{(2j+1)}_{lm},
\end{gather*}
with coef\/f\/icients
\begin{gather*}
F^{(2j+1)}_{lm}=\frac{\tr\big[\hat{Y}^{(2j+1)\dagger}_{lm} \hat{F}^{(2j+1)}\big]}
{\tr\hat{Y}^{(2j+1)\dagger}_{lm}\hat{Y}^{(2j+1)}_{lm}}.
\end{gather*}
Fuzzy harmonics are def\/ined as the symbols of spherical harmonics operators over coherent states (cf.\
our previous def\/inition~\eqref{fuzzyha})
\begin{gather}
{}_{2j+1}\langle\vartheta, \varphi|\hat{Y}^{({2j+1})}_{lm}|\vartheta,\varphi\rangle_{2j+1}=
Y^{({2j+1})}_{lm}(\vartheta,\varphi).
\label{fuzzyharm}
\end{gather}
They form a~basis in the noncommutative algebra $C(S^{2})$.
We have indeed
\begin{gather*}
{f}^{(2j+1)}= \sum\limits_{l=0}^{j}\sum\limits_{m=-l}^{l}
F^{(2j+1)}_{lm} {Y}^{(2j+1)}_{lm}.
\end{gather*}

A Weyl map $\Omega_{2j+1}:C(S^2)\to\mathbb M_{2j+1}(\mathbb C)$ is def\/ined by simply mapping spherical harmonics into
spherical harmonics operators,
\begin{gather*}
\Omega_{2j+1}\left(Y_{lm} \left(\vartheta,\varphi\right)\right)=
\begin{cases}
\hat{Y}^{({2j+1})}_{lm}, &l\leq j,
\\
0, &l>j,
\end{cases}
\end{gather*}
and extending the map by linearity.
One can def\/ine the adjoint map as
\begin{gather*}
\Omega_{2j+1}^{-1}\big(\hat{Y}^{(2j+1)}_{lm}\big)=Y^{({2j+1})}_{lm} (\vartheta,\varphi).
\end{gather*}
Using the Berezin symbol~\eqref{berezinsymsph} it is possible to identify the adjoint map:
\begin{gather*}
\Omega_{2j+1}^\dagger\big(\hat F^{({2j+1})}\big)(\vartheta,\varphi)=\bra{\vartheta,\varphi}\hat
F^{({2j+1})}\ket{\vartheta,\varphi},
\end{gather*}
for a~matrix $\hat F^{{2j+1}}$.
The two maps are one the adjoint of the other in the sense that
\begin{gather*}
\langle{\Omega_{2j+1}}(f), \hat G^{({2j+1})}\rangle_{2j+1}=\langle f, \Omega^\dagger_{2j+1}\big(\hat
G^{({2j+1})}\big)\rangle_{L^2(S^2)}
\end{gather*}
for all $f\in C(S^2)$ and all $\hat G^{({2j+1})}\in\mathbb M_{2j+1}(\mathbb C)$.
The f\/irst scalar product is taken in the f\/inite-dimensional Hilbert space $\mathbb C^{2j+1}$, while the second one is
taken in $L^2(S^2)$.

The f\/inite-dimensional matrix algebra is mapped by $\Omega_{2j+1}$ into a~subspace of the inf\/inite-dimensional space of
functions on $S^2$.
Restricting the functions of the sphere on this subspace makes $\Omega_{2j+1}^\dagger=\Omega_{2j+1}^{-1}$.
This subspace is not an algebra under the usual commutative product of functions, but it is a~\emph{noncommutative}
algebra under the $\star$-product def\/ined as usual~by
\begin{gather*}
(f*g)(\vartheta,\varphi)=\Omega_{2j+1}^{-1}(\Omega_{2j+1}(f)\Omega_{2j+1}(g)).
\end{gather*}
This $*$ product is given by the symbol of the product of two fuzzy
harmonics~\cite{ChuMadoreSteinacker,Hoppethesis,IKTW}, which can be obtained in term of $6j$-symbols~\cite{Varsalovich}
\begin{gather}
\hat{Y}^{(2j+1)}_{l^{\prime}m^{\prime}} \hat{Y}^{(2j+1)}_{l^{\prime\prime}m^{\prime\prime}} =
\sum\limits_{l=0}^j(-1)^{2j+l}\sqrt{{{\frac{(2l^{\prime}+1)
\left(2l^{\prime\prime}+1\right)(2j-l)(2j+l'+1)(2j+l''+1)} {4\pi(2j+l+1)(2j+l'+1)(2j+l''+1)}}}}
\nonumber
\\
\phantom{\hat{Y}^{(2j+1)}_{l^{\prime}m^{\prime}} \hat{Y}^{(2j+1)}_{l^{\prime\prime}m^{\prime\prime}} =}
\times\left\{
\begin{matrix}
{l^{\prime}} & {l^{\prime\prime}} & {\scriptstyle l}
\\
{j} & {j} & {j}
\end{matrix}
\right\} C^{lm}_{l^{\prime}m^{\prime} l^{\prime\prime}m^{\prime\prime}} \hat{Y}^{(2j+1)}_{lm}.
\label{prodYcomm}
\end{gather}

The fuzzy harmonics def\/ined in~\eqref{fuzzyharm} are the eigenvectors of the \emph{fuzzy Laplacian}.
The natural inf\/initesimal action of ${\rm SU}(2)$ on $\mathbb M_{2j+1}(\mathbb C)$ is given by the adjoint action $\hat
F^{({2j+1})}\mapsto [J_i,\hat F^{({2j+1})}]$ of the generators $J_i$ in the $(2j+1)$-dimensional representation.
With these three derivations we def\/ine the fuzzy Laplacian by the symbol of the operator
\begin{gather*}
\nabla^{2}  : \ \ \mathbb{M}_{{2j+1}}(\mathbb C) \mapsto \mathbb{M}_{{2j+1}}(\mathbb C),
\\
\nabla^{2}f^{({2j+1})} = {}_{2j+1}\langle\vartheta, \varphi| \nabla^{2}\hat{F}^{(2j+1)}|
\vartheta,\varphi\rangle{}_{2j+1}
 = {}_{2j+1}\langle\vartheta, \varphi|\sum\limits_{i=1}^3\big[{J}_{i},\big[{L}_{i},\hat{F}^{(2j+1)}\big]\big]| \vartheta,\varphi\rangle{}_{2j+1},
\end{gather*}
where, with an abuse of notation, we use the same symbol for the operator acting on $\mathbb{M}_{{2j+1}}(\mathbb C)$ and
the fuzzy Laplacian, which properly acts on the algebra of functions on the sphere $C(S^2)$.
Its spectrum consists of eigenvalues $l(l+1)$, where $l=0,\ldots, {2j+1}$, and every eigenvalue has a~multiplicity $2l+1$.
The spectrum of the fuzzy Laplacian thus coincides up to order ${2j+1}$ with that of its continuum counterpart.

As in the case of the fuzzy torus described earlier the fuzzy sphere converges to the usual sphere.
Using properties of the $6j$-symbols in the product~\eqref{prodYcomm} one can argue that the $j\to\infty$ limit of this
product reproduces the standard product of spherical harmonics.
This gives a~naive way to see that in the limit the fuzzy sphere algebra becomes the algebra of functions on~$S^2$.

In the sphere case there are rigorous proofs that this happens in a~precise mathematical sense~\cite{Rieffel0108005}.
The proof is based on the fact that the fuzzy sphere structure gives the algebra of matrices a~metric structure of
a~distance among states.
It is possible also to prove that the distance between the coherent states def\/ined above converges to the metric
distance on the sphere~\cite{DAndreaLizziVarilly}.
Def\/ining a~distance among metric spaces makes it possible to show that the distance between the fuzzy spheres and the
ordinary sphere goes to zero as $j\to\infty$.

The fuzzy sphere as a~matrix model has been studied extensively as a~matrix model of f\/ield theories, see for example the
reviews~\cite{Abe, Panero}.

\subsection{The fuzzy disc}\label{Section8.3}

We have considered in Section~\ref{Section4.2} the matrix basis for the Wick--Voros product on the plane.
Let un now truncate the algebra ${\mathbb R}^2_\theta$ with the projector
\begin{gather*}
\hat P^{(N)}_\theta=\sum\limits_{n=0}^N |n\rangle\langle n|.
\end{gather*}
The symbol of this operator is the function
\begin{gather}
P^{(N)}_\theta(r,\varphi)=\sum\limits_{n=0}^N \langle z|n\rangle\langle n|z\rangle={\mathrm
e}^{\frac{r^2}\theta}\sum\limits_{n=0}^N \frac{r^{2n}}{\theta^n
n!}=\frac{\Gamma\left(N+1,r^{2}/\theta\right)}{\Gamma\left(N+1\right)},
\label{projectorfunction}
\end{gather}
where
we use the usual polar decomposition $z=r{\mathrm e}^{{\mathrm i}\varphi}$.
By construction $P^{(N)}_\theta\star_V P^{(N)}_\theta=P^{(N)}_\theta$.

The disc~\cite{fuzzydisc2,fuzzydisc1,fuzzydisc3,fuzzydisc4} {(see also~\cite{BGK})} is recovered considering the
simultaneous limit
\begin{gather}
N\to\infty;
\qquad
\theta\to 0
\qquad
\text{with}
\quad
N\theta=R^2,
\label{Nthetalimit}
\end{gather}
where~$R$ will be the radius of the disc.
In the following we take $R^2=1$ to simplify notations.
In this case the limit~\eqref{projectorfunction} can be performed using known properties of incomplete Gamma functions
to obtain
\begin{gather*}
P^{(N )}_{\theta}
\
\rightarrow
\
\begin{cases}
1 & r<1,
\\
1/2 & r=1,
\\
0 & r> 1.
\end{cases}
\end{gather*}
In other words, the symbols of the projector $\hat P^{(N)}_{\theta}$ is an approximation of the
characteristic function of the disc, and converges to in the limit~\eqref{Nthetalimit}.
This suggests to consider, in analogy with the fuzzy sphere, a~f\/inite matrix algebra, $\hat A^N_\theta$ (or rather
a~sequence of algebras), whose symbols are functions with support on a~disc.
The fuzzy disc is thus def\/ined as the sequence of subalgebras $\mathcal{A}^N_\theta$,
\begin{gather*}
\mathcal{A}^N_\theta= P^{(N )}_{\theta}\star {\mathbb R}^2_\theta\star P^{(N )}_{\theta},
\end{gather*}
with ${\mathbb R}^2_\theta$ the Wick Voros algebra on the plane.

{A dual view, i.e.~taking the projector $\mathbb I-P^{(N)}$ gives a~Moyal plane with a~``defect''~\cite{PinzulStern},
spherical wells have also been considered~\cite{SCGV}}.
In order to do this we consider f\/irst the Laplacian basis of functions for the disc with Dirichlet boundary conditions.
For the Laplacian on the disc all eigenvalues are negative, their modules~$\lambda$ are obtained solving for the zeroes
of the Bessel functions:
\begin{gather*}
J_{n}\big(\sqrt{\lambda}\big)=0.
\end{gather*}
They are doubly degenerate for~$n$ non zero, in which case they are simply degenerate.
We label them $\lambda_{n,k}$ where~$k$ indicates that it is the $k^{th}$ zero of the function.
The eigenfunctions are:
\begin{gather*}
\Phi_{n,k}={\rm e}^{in\varphi} \left(\frac{\sqrt{\lambda_{\left|n\right|,k}}r}{2}\right)^{|n|} \sum\limits_{s=0}^{\infty}
\frac{\left(-\lambda_{|n|,k}\right)^{s}}{s!\left(|n|+s\right)!} \left(\frac{r}{2}\right)^{2s}=
{\rm e}^{in\varphi}J_{|n|}\left(\sqrt{\lambda_{\left|n\right|,k}}r\right).
\end{gather*}

\begin{figure}[htb] \centering
\includegraphics[width=0.32\linewidth]{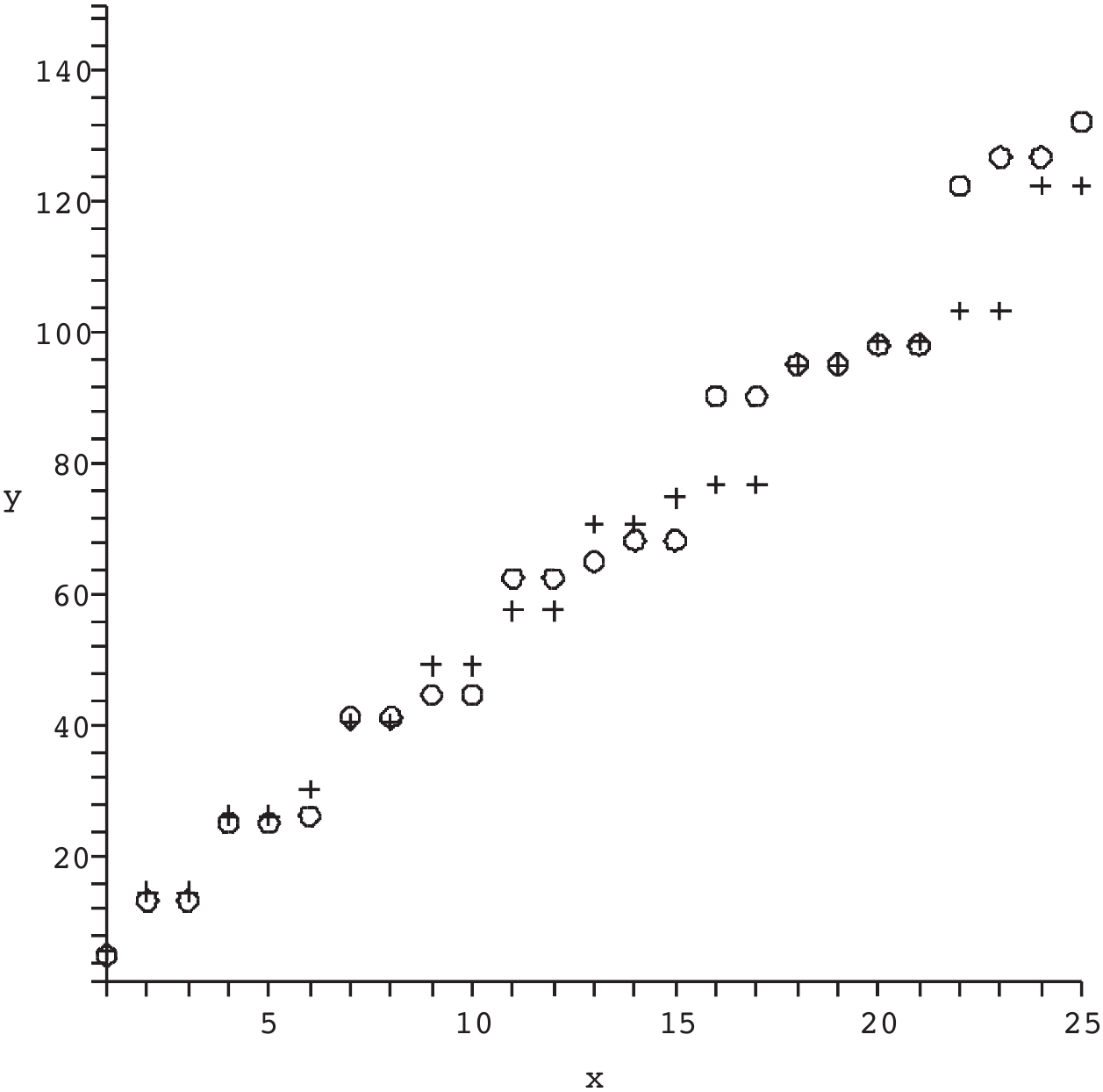}
\includegraphics[width=0.32\linewidth]{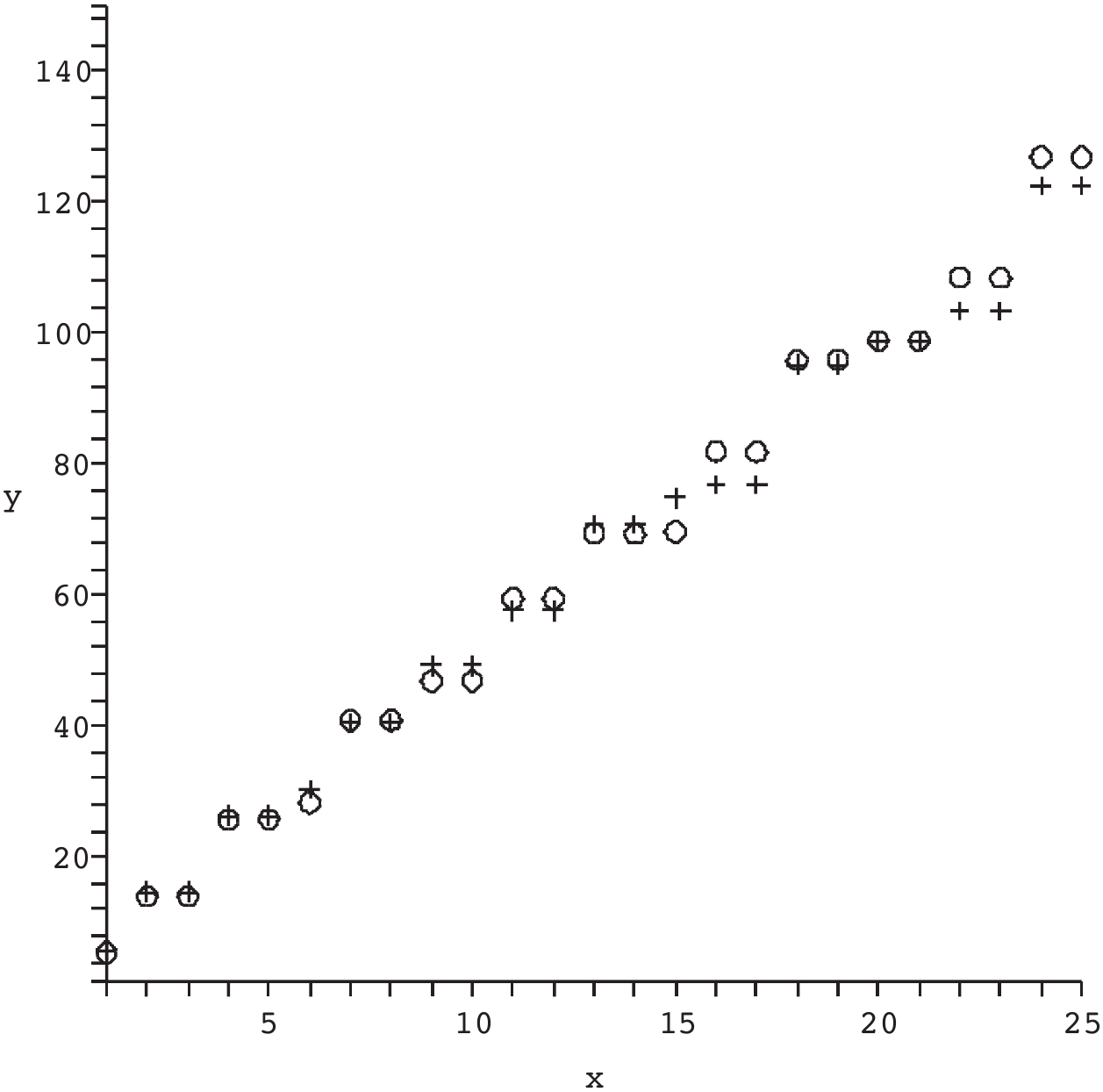}
\includegraphics[width=0.32\linewidth]{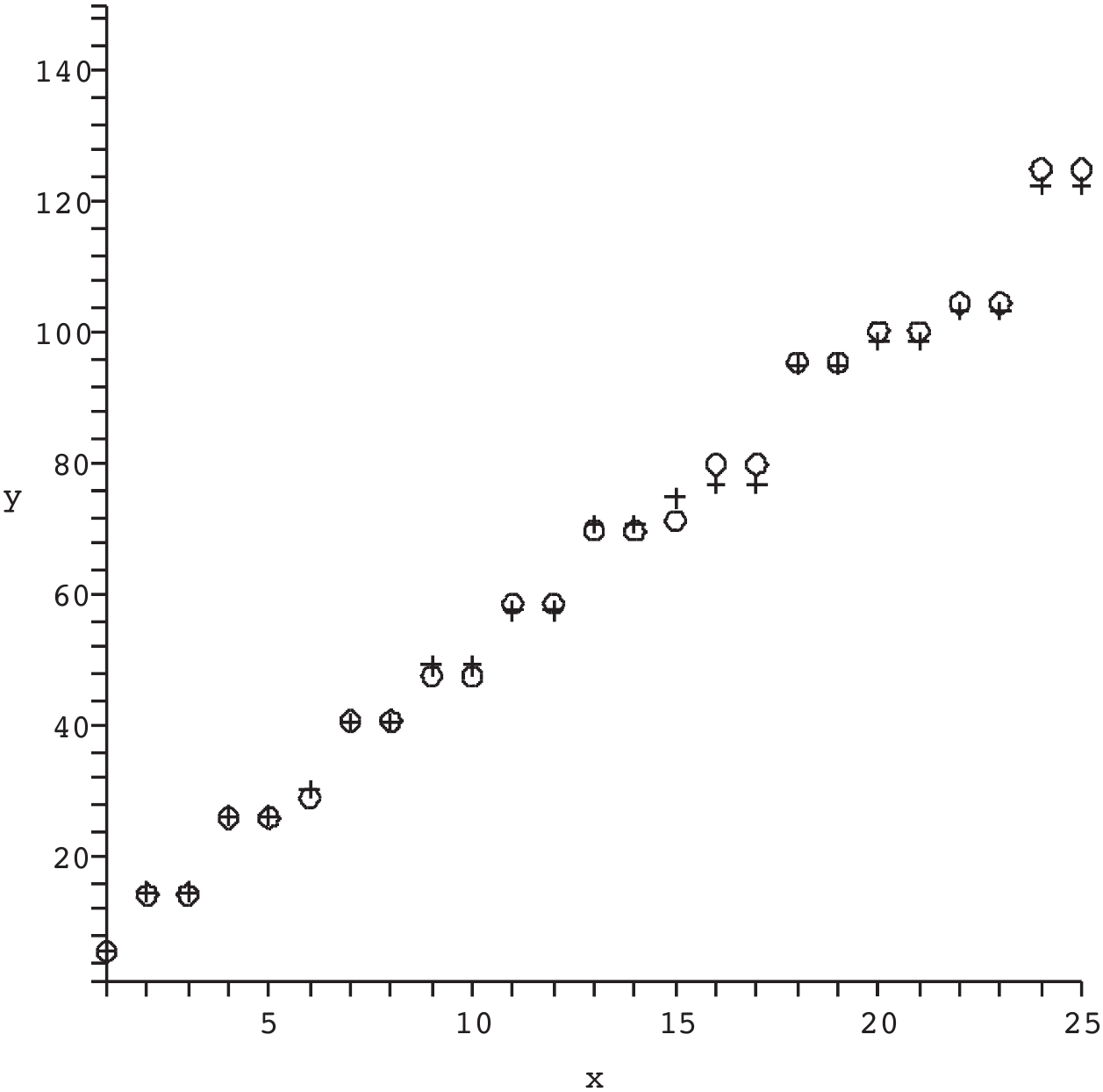}

\caption{Comparison of the f\/irst eigenvalues of the fuzzy Laplacian (circles) with those of the continuum Laplacian
(crosses) on the domain of functions with Dirichlet homogeneous boundary conditions.
The orders of truncation are $N=10,20,30$.}
\label{fuzzydrum}
\end{figure}

Because of relation~\eqref{derivcommz} it is possible to express the Laplacian in terms of inner derivations, and
therefore, after the projection, express it as an automorphism of the algebra of matrices, and f\/ind the eigenvectors of
it.
From the exact expression on the plane:
\begin{gather*}
\nabla^2 f(\bar z,z)=4\partial_{\bar{z}}\partial_{z}f =\frac{4}{\theta^{2}}[z,[f,\bar z]_\star]_\star
\end{gather*}
it is possible to def\/ine, in each $\mathcal{A}^{(N)}_{\theta}$:
\begin{gather*}
\nabla^2_{(N )}  f^{(N )}_{\theta} =\langle z| \nabla^2_{(N )} \hat
f^{(N )}_{\theta} |z\rangle \equiv\frac{4}{\theta^{2}}\langle z|\hat{P}^{(N )}_{\theta}
\big[\hat{a},\big[\hat{P}^{(N )}_{\theta}
\hat{f}\hat{P}^{(N )}_{\theta},\hat{a}^{\dagger}\big]\big] \hat{P}^{(N )}_{\theta}|z\rangle.
\end{gather*}
The spectrum of this \emph{fuzzy Laplacian} is of course f\/inite, but it is possible to see from Fig.~\ref{fuzzydrum}
that it approaches the spectrum of the continuous Laplacian as~$N$ increases (with $N\theta=1$).

\begin{figure}[htb] \centering
\includegraphics[width=0.3\linewidth]{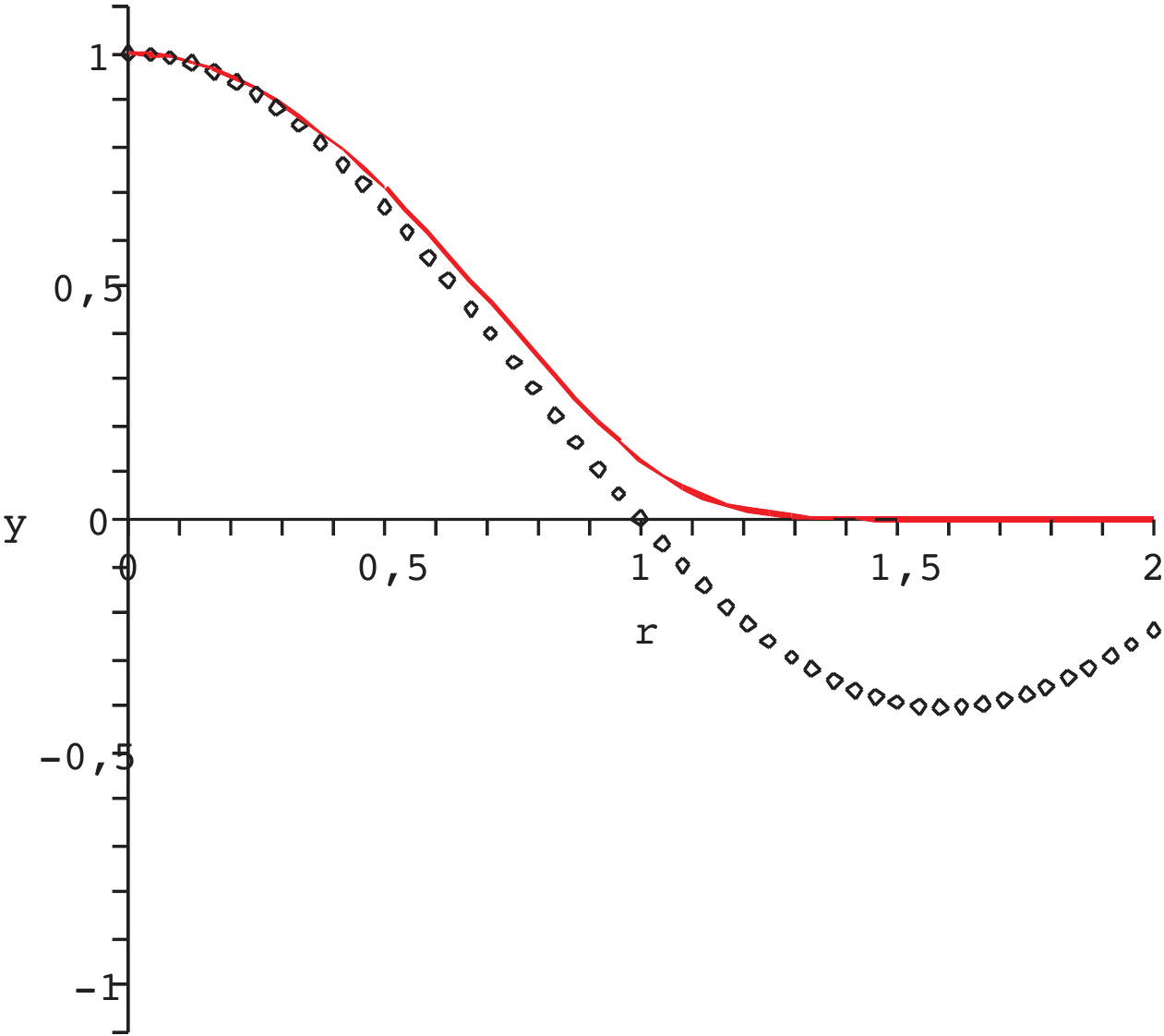}
\includegraphics[width=0.3\linewidth]{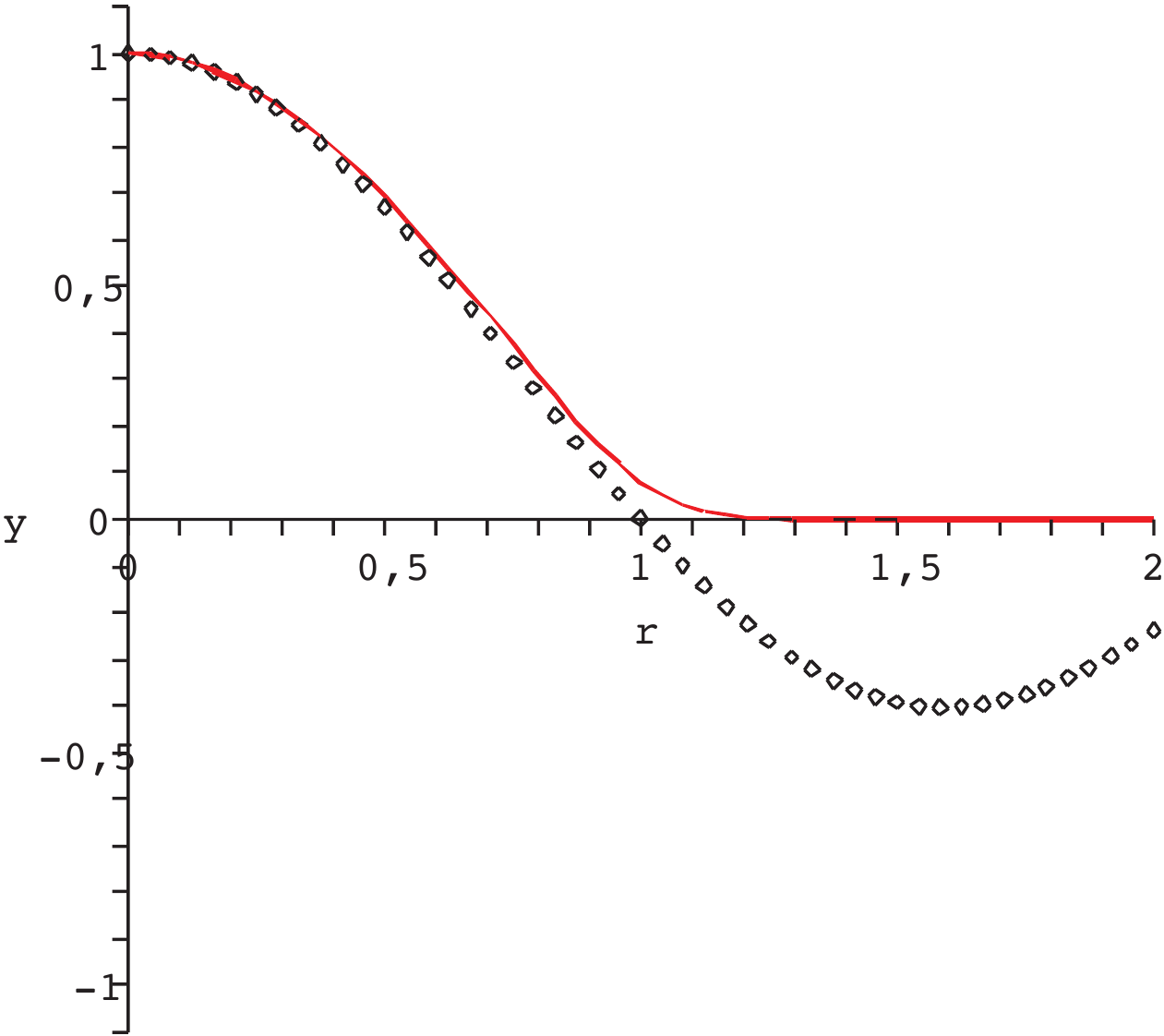}
\includegraphics[width=0.3\linewidth]{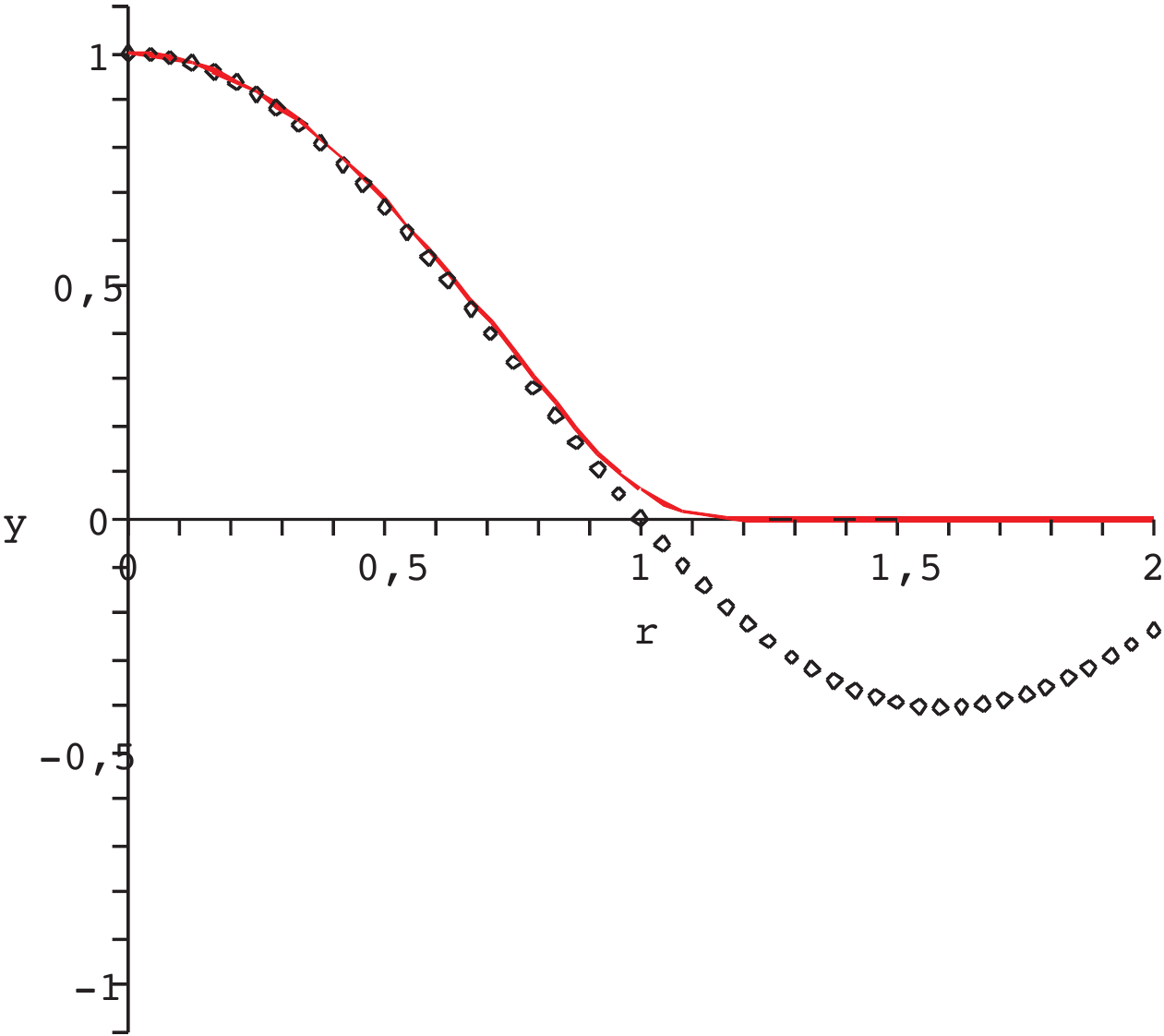}

\caption{Comparison of the radial shape for the symbol $\Phi^{(N)}_{0,1}(r,\varphi)$ (continuum
line), the symbol of the eigenmatrix of the fuzzy Laplacian for $N=10,20,30$, with $\Phi_{0,1}(r,\varphi)$.}
\label{zero_uno}
\end{figure}

The fuzzy Laplacian is an automorphism of the algebra on $n\times n$ matrices.
In analogy with the fuzzy harmonics described earlier we call the radial part of its eigenoperators {Fuzzy Bessel
operators}.
Their symbols, the \emph{Fuzzy Bessel functions}, form a~basis for the fuzzy disc and approximate well the actual Bessel
functions, as can be seen from Fig.~\ref{zero_uno}.

Likewise it is possible~\cite{KobayashiAsakawa}, using the phase operator and phase states known in quantum optics, to
have \emph{fuzzy angles}, i.e.~some states concentrated in a~small angular region of the disc.
Field theories on the fuzzy disc have been studied in~\cite{FFPV, spisso}.

\section{Matrix models and the emergence of gravity} \label{Section9}

Matrix models have a~long and distinguished history, especially in string theory~\cite{BFSS, DijkgraafVerlinde, IKKT},
in this review we would like to discuss brief\/ly how the discrete basis of the noncommutative products describe earlier
gives rise to a~matrix model in which gravity is contemplated as an emergent phenomenon, much like the emergent gravity
of Sakharov~\cite{Sakharov}.
Here by emergent gravity we really mean the emergence of f\/ields moving in a~curved background.
This is a~more modest goal than having the metric degrees of freedom emerging as quantized f\/ields.
This would be tantamount to have a~full theory of quantum gravity.
And as is known, we are not yet there~\dots.

Since for the kind of products we are considering derivations are inner automorphisms of the algebra\footnote{One would
have to def\/ine precisely which algebra is being considered, since for example the coordinate functions do not belong to
the algebra of Schwarzian functions with the Moyal product.
They however belong to the multiplier algebra.} since they can be expressed by a~commutator:
\begin{gather}
\frac\partial{\partial x_\mu} f={\mathrm i}\theta_{\mu\nu}^{-1}[x^\nu,f]_*.
\label{derivcomm}
\end{gather}
If one considers a~U(1) gauge theory on this space, with unitary transformations given by star unitary elements
$U*U^\dagger=\mathbb I$, the action invariant for the transformation $F\to U*F*U^\dagger$ is given~by
\begin{gather*}
S=-\int {\mathrm d} x \frac14 F*F.
\end{gather*}
One can def\/ine a~covariant derivative
\begin{gather*}
D_\mu f =\partial_\mu f - {\mathrm i} [f,A_\mu]_* ={\mathrm i}\theta^{-1}_{\mu\nu}[X^\mu,f]_*
\end{gather*}
and
\begin{gather*}
F^{\mu\nu}=[D^\mu,D^\nu]_\star=[X^\mu,X^\nu]_\star + \theta^{\mu\nu}.
\end{gather*}
The connection between commutator with the coordinates and derivatives~\eqref{derivcomm}, suggest~\cite{MSSW} the
def\/inition of \emph{covariant coordinates}
\begin{gather*}
X^\mu=x^\mu + \theta^{\mu\nu} A_\nu
\end{gather*}
and consequently
\begin{gather*}
D_\mu f ={\mathrm i}\theta^{-1}_{\mu\nu}[X^\mu,f]_\star=\partial_\mu f - {\mathrm i} [f,A_\mu]_\star.
\end{gather*}
Therefore we have
\begin{gather*}
F^{\mu\nu}=[D^\mu,D^\nu]_\star=[X^\mu,X^\nu]_\star + \theta^{\mu\nu}.
\end{gather*}
The constant~$\theta$ can be reabsorbed by a~f\/ield redef\/inition and the action is the square of this quantity,
integrated over spacetime.

The action can therefore be rewritten, in the matrix basis as
\begin{gather}
S=-\frac1{4g}\tr [X^\mu,X^\nu][X^{\mu'},X^{\nu'}]g_{\mu\mu'}g_{\nu\nu'},
\label{bosonicaction}
\end{gather}
where the~$X$'s are operators (matrices) and the metric $g_{\mu\mu'}$ is the \emph{flat} Minkowski (or Euclidean)
metric.

We now brief\/ly remind how gravity emerges from this model~\cite{Steinackeroriginal}.
The equations of motion corresponding to the action~\eqref{bosonicaction}:
\begin{gather*}
[X^\mu,[X^\nu,X^{\mu'}]]g_{\mu\mu'}=0.
\end{gather*}
These equations have dif\/ferent solutions, which we call vacua.
One solution in particular corresponds to the star product generated by~\eqref{xcomm}.
We call the matrixes correspondding to this particular solution $X_0$, hence
\begin{gather}
[X_0^\mu,X_0^\nu]={\mathrm i}\theta^{\mu\nu}.
\label{Moyalmatrixvacuum}
\end{gather}
Note that the relation can only be valid if the~$X$'s are inf\/inite matrices corresponding to non bounded operators.
Fluctuations around the $X_0$'s will give a~generalized commutation relation
\begin{gather*}
[X^\mu,X^\nu]={\mathrm i}\theta(X),
\end{gather*}
where we have def\/ined the matrices $X^\mu=X_0^\mu+A^\mu$ in analogy with the covariant coordinates.

An important result obtained in~\cite{Steinackeroriginal} (see also~\cite{Yang}) is obtained if one couples the theory
to a~scaler f\/ield~$\Sigma$.
At this stage the meaning and origin of this f\/ield is yet undetermined, it is a~f\/ield which couples to the
noncommutative space time.
The free action of this f\/ield, using the fact that the derivative are expressed as commutators with the coordinates, is:
\begin{gather*}
\tr [X^\mu,\Sigma][X^\nu,\Sigma]g_{\mu\nu}\sim\int{\mathrm d} x (D_{\mu'}\Sigma)
(D_{\nu'}\Sigma)\theta^{\mu\mu'}\theta^{\nu\nu'}g_{\mu\nu}= \int{\mathrm d} x (D_{\mu}\Sigma)(D_{\nu}\Sigma)G^{\mu\nu}.
\end{gather*}
But it easy to recognize the fact that this is the action of f\/ield moving in a~non f\/lat background described by the
metric
\begin{gather*}
G^{\mu\nu}(x)=\theta^{\mu\mu'}\theta^{\nu\nu'}g_{\mu'\nu'}.
\end{gather*}
The mere fact that the f\/ield was moving in a~noncommutative space described by the matrix model has induce a~curved
background, so that gravity appears as an \emph{emergent} phenomenon.

The vacuum~\eqref{Moyalmatrixvacuum} is not the only one.
One can consider alternative vacua which have an invariance for some group.
For example
\begin{gather*}
\bar X^\mu_0=X_0^\mu\otimes\mathbb I_{n}.
\end{gather*}
In this case the theory has an internal space, and a~noncommutative ${\rm U}(n)$ symmetry.
However the ${\rm U}(1)$ degree of freedom of the theory is the described above, which couple gravitationally.
One can separate the trace part $A_0$ form the the traceless generators of ${\rm SU}(n)$, considering as f\/luctuations
\begin{gather*}
\bar X=\bar X_0+A_0+A_\alpha\lambda_\alpha.
\end{gather*}
Several models can be constructed based on these matrix models.
Extra dimensions can appear in the form of fuzzy spheres~\cite{AtanasiosHaroldGeorge1} and it is possible to have models
which start having also characteristics of the standard model~\cite{AtanasiosHaroldGeorge2,GrosseLizziSteinacker,
SteinackerZahn}.

\subsection*{Acknowledgements}

We were partially supported by UniNA and Compagnia di San Paolo under the grant ``Programma STAR 2013''.
F.~Lizzi acknowledges support by CUR Generalitat de Catalunya under project FPA2010-20807.

\pdfbookmark[1]{References}{ref}
\LastPageEnding

\end{document}